\pdfoutput=1
\documentclass[9.5pt,journal,onecolumn]{IEEEtran}
\IEEEoverridecommandlockouts

\usepackage[utf8]{inputenc}
\DeclareUnicodeCharacter{03B2}{\ensuremath{\beta}}

\usepackage{amsmath}
\usepackage{amsthm}
\usepackage{amssymb}
\usepackage{comment}
\usepackage{graphicx}
\usepackage{gensymb}
\usepackage{textcomp}
\usepackage{caption}
\usepackage{algorithm}
\usepackage{algpseudocode}
\usepackage{csvsimple}
\usepackage{array}
\usepackage{flushend}
\usepackage{balance}
\usepackage{multirow}
\usepackage{mathtools}
\usepackage{bm}
\usepackage{xcolor}
\usepackage{soul}
\usepackage{diagbox}
\usepackage{tablefootnote}

\usepackage[caption=false, font=footnotesize]{subfig}
\usepackage[bookmarks=false]{hyperref}
\hypersetup{
    colorlinks = true,
    citecolor  = blue,
    linkcolor  = blue,
    urlcolor   = blue,
}

\DeclarePairedDelimiter\floor{\lfloor}{\rfloor}
\DeclarePairedDelimiter\abs{\lvert}{\rvert}
\newcommand{\Expect}{{\rm I\kern-.3em E}}
\newcommand{\Enc}{\operatornamewithlimits{Enc}}

\newcommand{\LFSR}{\operatornamewithlimits{LFSR}}

\newcommand{\Gen}{\operatornamewithlimits{Gen}}
\newcommand{\Dec}{\operatornamewithlimits{Dec}}

\providecommand{\innerprod}[1]{\langle#1\rangle}

\newtheorem{theorem}{Theorem}
\newtheorem{definition}{Definition}
\newtheorem{lemma}{Lemma}

\usepackage{tikz}
\usepackage{textcomp}
\usepackage{hyperref}
\usepackage{lipsum}

\newcommand\copyrighttext{%
  \footnotesize \textcopyright 2023 IEEE. Personal use of this material is permitted.
  Permission from IEEE must be obtained for all other uses, in any current or future
  media, including reprinting/republishing this material for advertising or promotional
  purposes, creating new collective works, for resale or redistribution to servers or
  lists, or reuse of any copyrighted component of this work in other works.
  DOI: \href{https://ieeexplore.ieee.org/abstract/document/9900128}{10.1109/TC.2022.3207119}}
\newcommand\copyrightnotice{%
\begin{tikzpicture}[remember picture,overlay]
\node[anchor=south,yshift=10pt] at (current page.south) {\fbox{\parbox{\dimexpr\textwidth-\fboxsep-\fboxrule\relax}{\copyrighttext}}};
\end{tikzpicture}%
}

\begin{document}
\setstcolor{red}

\title{A Provably Secure Strong PUF based on LWE: Construction and Implementation}

\author{Xiaodan Xi\textsuperscript{*}, Ge Li\textsuperscript{*}, Ye Wang, Yeonsoo Jeon, and Michael Orshansky

\thanks{The authors are with the Department of Electrical and Computer Engineering, The University of Texas at Austin, Austin, TX, 78712, USA.
E-mail: paul.xiaodan@utexas.edu; lige@utexas.edu; lhywang@utexas.edu; yeonsoo@utexas.edu;  orshansky@utexas.edu}

\thanks{*Xiaodan Xi and Ge Li contributed equally to this work.}

}

\maketitle
\copyrightnotice

\begin{abstract}
We construct a strong PUF with provable security against ML attacks on both classical and quantum computers. The security is guaranteed by the cryptographic hardness of learning decryption functions of public-key cryptosystems, and the hardness of the learning-with-errors (LWE) problem defined on integer lattices. We call our construction the lattice PUF.

We construct lattice PUF with a physically obfuscated key and an LWE decryption function block. To allow deployments in different scenarios, we demonstrate designs with different latency-area trade-offs. A compact design uses a highly serialized LFSR and LWE decryption function, while a latency-optimized design uses an unrolled LFSR and a parallel datapath.

We prototype lattice PUF designs with $2^{136}$ challenge-response pairs (CRPs) on a Spartan 6 FPGA. In addition to theoretical security guarantee, we evaluate empirical resistance to the various leading ML techniques: the prediction error remains above $49.76\%$ after $1$ million training CRPs. The resource-efficient design requires only $45$ slices for the PUF logic proper, and $351$ slices for a fuzzy extractor. The latency-optimized design achieves a $148X$ reduction in latency, at a $10X$ increase in PUF hardware utilization. 
The mean uniformity of PUF responses is $49.98\%$, the mean uniqueness is $50.00\%$, and the mean reliability is $1.26\%$.

\end{abstract}


\IEEEpeerreviewmaketitle

\section{Introduction}
\label{sec:intro}
\IEEEPARstart{S}{ilicon} physical unclonable functions (PUFs) are security primitives commonly adopted for device identification, authentication, and cryptographic key generation \cite{suh2007physical}.
A PUF exploits the inherent randomness of CMOS technology to generate an output response for a given input challenge. 
Weak PUFs, which are also called physically obfuscated keys (POKs) \cite{herder2017trapdoor}, have a limited challenge-response pair (CRP) space. 
In contrast, strong PUFs supply an exponentially large number of CRPs.

The security of a strong PUF requires the associated CRPs to be unpredictable: given a certain set of known CRPs, it should be hard to predict the unobserved CRPs, even with the most powerful machine learning (ML) based modeling attacks. 
Engineering an ML resistant strong PUF with low hardware cost has been challenging. 

Variants of the original arbiter PUF (APUF) with stronger modeling attack resilience, including bistable ring PUF and feed-forward APUF, have also been broken via  ML attacks \cite{schuster2014evaluation, ruhrmair2010modeling}. 
The recent interpose PUF (IPUF) \cite{nguyen2019interpose} proposal is claimed to have provable ML resistance. 
Its security is rigorously reduced to the assumption that XOR APUFs are hard to model. 
Unfortunately, \cite{DBLP:journals/iacr/SantikellurBC19} demonstrates that both XOR APUFs and IPUFs are actually vulnerable to deep-learning-based modeling attacks.

By exploiting transistor-level intrinsic nonlinearity, some strong PUFs \cite{kumar2014design, zhuang2019strong} exhibit empirically-demonstrated resistance to multiple ML algorithms. 
Empirical demonstrations of ML resistance are not fully satisfactory since they can never rule out the possibility of other more effective ML algorithms. 

The controlled PUF \cite{gassend2008controlled} uses cryptographic primitives, such as hash functions, to ensure ML resistance. 
However, hardware implementation of hash functions usually requires a large area.
Strong PUF constructions using established cryptographic ciphers, such as AES \cite{bhargava2014efficient}, obtain ML resistance from the security of the ciphers, but have similar area overheads. 
\cite{fuller2013computational} has utilized lattice-based problems, including learning-parity-with-noise (LPN) and learning-with-errors (LWE), to realize computationally-secure fuzzy extractors (CFEs).
\footnote{A CFE guarantees absence of information leakage from publicly shared helper data via computational hardness in contrast to conventional FEs that need to limit their information-theoretic entropy leakage.} 
As a byproduct, the CFE-based strong PUF is constructed in \cite{herder2017trapdoor,jin2017fpga}.
However, as we discuss later, and pointed out in \cite{herder2017trapdoor,jin2017fpga}, a direct strong PUF implementation using CFE in \cite{fuller2013computational} may be vulnerable to attacks that compromise the hardness of LPN or LWE problems. 
\cite{herder2017trapdoor} and \cite{jin2017fpga} introduce a cryptographic hash function to hide the original CRPs to fix the vulnerability and achieve ML resistance, but this increases hardware cost.

\emph{In this paper, we propose a strong PUF that is secure against ML attacks with both classical and quantum computers.} \cite{wang2020lattice} 
As a formal framework to define ML resistance, we adopt the probably approximately correct (PAC) theory of learning \cite{mohri2012foundations}. 
Specifically, the PAC non-learnability of a decryption function implies that with a polynomial number of samples, with high probability, \emph{it is not possible to learn the function accurately in a polynomial time by any means.}
The main insight, which allows us to build such a novel strong PUF, is our reliance on the earlier proof that PAC-learning a decryption function of a semantically secure public-key cryptosystem entails breaking that cryptosystem \cite{kearns1994cryptographic, klivans2006cryptographic}.
We develop a PUF for which the task of modeling is equivalent to PAC-learning the decryption function of a LWE public-key cryptosystem.
The security of LWE cryptosystems relies on the hardness of LWE problem that ultimately is reduced to the hardness of several problems on lattices \cite{regev2009lattices}. 
Notably, LWE is believed to be secure against both classical and quantum computers.
Due to the intrinsic connection between the proposed PUF and the lattice cryptography we call our construction the \textbf{lattice PUF}.

A lattice PUF is realized by two major components: a POK and an LWE decryption function. 
The POK is needed for silicon-intrinsic key generation, with all the well-known security benefits of PUF-based key generation. 
A non-volatile memory based key storage mechanism in place of the POK is an alternative approach, but is not able to fulfill the strong PUF definition with reduced security against physical key-extraction attacks.
The core module of the lattice PUF is the LWE decryption function.
It generates a response (plaintext) for each submitted challenge (ciphertext). 
By itself, this block represents a keyed LWE-based hash function with restricted inputs (which follow the ciphertext distribution). 
The restriction on inputs permits a more compact design compared to other known LWE-based pseudorandom functions, e.g. \cite{brenner2014fpga}.
The design is carefully crafted considering trade-offs between various aspects.
We use the total number of operations needed to learn a model of the PUF, as a measure of ML security \cite{lindner2011better, micciancio2009lattice, albrecht2015concrete}. 
Using this estimator \cite{albrecht2015concrete}, we say that a PUF has $k$-bit ML resistance if a successful ML attack requires $2^k$ operations. 
Our implementation of the LWE decryption function is configured to achieve a $128$-bit ML resistance.

Our reliance on LWE to formally guarantee resistance to ML attacks is novel.
We highlight a critical difference between our work and recent work \cite{herder2017trapdoor,jin2017fpga}.
In short, the input-output mapping between the PUF of \cite{herder2017trapdoor,jin2017fpga} and the underlying LPN or LWE cryptosystem can be briefly summarized as follows: 
 challenges $\Longleftrightarrow$ public keys and responses $\Longleftrightarrow$  private keys. In contrast, the mapping of our construction is: challenges $\Longleftrightarrow$ ciphertexts and responses $\Longleftrightarrow$ decrypted plaintexts.

Although both constructions utilize lattice-based cryptosystems, they differ fundamentally in the root of their security guarantees.
The fundamental security property that \cite{fuller2013computational} rely upon is the \emph{computational hardness of recovering a private key from a public key in a public-key cryptosystem}. 
It turns out that, when building strong PUFs using public keys as challenges and private keys as responses, this security is insufficient. 
The vulnerability is due to the fact that the challenges are by definition publicly known \cite{fuller2013computational,herder2017trapdoor,jin2017fpga}, and an attacker can use multiple challenges (public keys) to recover the private key. 
This is only possible because multiple public keys are derived using a fixed (same) source of secret POK bits, embedded in the error term of LPN or LWE. 
As was shown in \cite{apon2017efficient}, the fact that multiple CRPs have shared error terms can be easily exploited, which allows a computationally-inexpensive algorithm for solving an LPN or LWE instance. 

\emph{In stark contrast, the proposed lattice PUF derives its security by directly exploiting a distinctly different property of public-key cryptosystems: the theoretically-proven guarantee that their decryption functions are not PAC learnable}. 
(As is shown later, this property stems from the semantic security of a cryptosystem coupled with the ease of generating multiple ciphertexts.) 
Lattice PUF does not have the vulnerability mentioned above since the publicly known challenges are ciphertexts and the security of the public-key cryptosystem guarantees that the fixed private key (the POK, in our case) cannot be recovered from ciphertexts.

Further, the construction of  \cite{herder2017trapdoor,jin2017fpga} is expensive since (1) recovering secrets from helper data requires solving a system of linear equations, and (2) implementing a hash function is costly.  
In contrast, our PUF is lightweight since it implements only the LWE decryption function rather than the expensive encryption function.

Since a PUF may be deployed in very different applications, it is critical to convert a theoretically sound construction into an efficient physical implementation. 
To allow practical deployments of our construction under different requirements, we investigate a series of lattice PUF designs targeting different scenarios. 
We first investigate a resource-efficient design, which is suitable for extremely resource-constrained environments, such as embedded/edge devices. 
Directly constructing a strong PUF from an LWE decryption function is not efficient since it requires $1288$ input challenge bits to produce $1$ response bit.
We develop an efficient design considering resource-constrained environments which significantly (by about 100X) reduces the communication cost associated with PUF response generation.
This is achieved by \emph{exploiting distributional relaxations allowed by recent work in space-efficient LWEs} \cite{galbraith2013space}.
This allows introducing a low-cost pseudo-random number generator (PRNG) implemented with a linear-feedback shift register (LFSR) requiring transmission of only a small external seed. 
Finally, while the primary goal of the paper is to construct a PUF that is secure against passive attacks, we also eliminate the risk of an active attack by using the technique in \cite{yu2016lockdown}: we embed the counter value from a self-incrementing counter into the challenge seed.
This eliminates the vulnerability to the active attack as the counter restricts the attacker's ability to fully control input challenges fed to the LWE decryption function.

We then investigate a latency-optimized design for performance-critical applications. 
Its target deployment platform has sufficient hardware resources and aims to guarantee fast response time. 
An example is a server in an industrial control environment in which authentication requests have a stringent latency budget. 
The resource-efficient design has a large response generation latency due to its highly serialized compute. 
We reduce latency by using parallelism. 
We propose two strategies: (1) parallelizing the LFSR and the LWE decryption function data-path (which produces a single response bit serially), and (2) parallelizing the LWE decryption function (which performs a single multiply-and-accumulate (MAC) sequentially). 
The first strategy is achieved via multiple instantiations of LFSR and LWE decryption blocks. 
The second strategy is achieved via multiple MAC units in the LWE decryption block, accompanied by an unrolled LFSR. 
The second strategy demonstrates better hardware efficiency, but its maximum parallelism is limited by the specific bit-serial LFSR implementation. 
The LFSR generator polynomial limits the maximum unrolling factor. 
We show that the optimal strategy is to initially use parallelization of the LWE decryption function, and then parallelize the LSFR and the LWE decryption datapath to achieve further latency improvement.

Our lattice PUF construction is different from direct authentication using an LWE public-key cryptosystem with a key generated from a POK in the following aspects. 
Though the ML resistance of lattice PUF is derived from LWE public-key cryptosystem, we do not assume an interface transferring only a public key. 
Instead, the enrollment phase in the PUF scenarios permits secret information transferred in a secure manner. 
Transferring a secret key in the enrollment phase enables efficient implementation. 
First, the computationally expensive discrete Gaussian sampling can happen on the server side. 
Second, it allows for the distributional relaxation and challenge compression in our construction as we demonstrate in Section \ref{sec:design}.
Third, the enrollment phase allows the server to store valid CRPs, then only valid CRPs can be sent to PUFs. 
Compared to the LWE public-key encryption scheme which generates ciphertexts dynamically on the fly, CRP pre-storage further permits lower latency in authentication.

Our lattice PUF construction achieves a CRP space size of $2^{136}$.
Statistical simulation shows the proposed lattice PUF has excellent uniformity, uniqueness, and reliability.
The mean (standard deviation) of uniformity is $49.98\%$ ($1.58\%$), and of inter-class Hamming distance (HD) is $50.00\%$ ($1.58\%$).
The mean BER (intra-class HD) is $1.26\%$.
Although we establish the security of lattice PUF theoretically, we perform empirical validation as a way to explore the possible attacks of distributional relaxations used, and as a way to give added overall confidence in the design. We test empirical ML resistance of lattice PUF with both conventional ML algorithms and more powerful deep neural networks (DNNs). The prediction error remains close to $50\%$ even after 1M training CRPs.
A $1280$-bit secret key is required by the proposed lattice PUF.
To reconstruct the stable POK bits efficiently, the construction uses a concatenated-code-based fuzzy extractor (FE). 
We also provide an end-to-end authentication scheme with the lattice PUF and FE.
With a $5\%$ bit error rate (BER) in SRAM cells, $6.36K$ cells are required to achieve a $10^{-6}$ failure rate in key reconstruction.
We implement the entire lattice PUF (except for raw SRAM cells) on a Spartan 6 FPGA.
The resource-efficient design consumes only $45$ slices. 
The concatenation-code-based FE takes $351$ slices. 
The latency-optimized design achieves a 148X latency reduction (to about 10 cycles per response bit generation), at the cost of a 10X increase in hardware utilization and 2.4X increase in communication.
\section{ML Resistance of LWE Decryption Functions  (Lattice PUF)}
\label{sec:lwe}

\subsection{ML Resistance as Hardness of PAC Learning}
A strong PUF can be modeled as a function $f:\mathcal{C}\rightarrow \mathcal{R}$ mapping from the challenge space $\mathcal{C}$ (usually $\{0,1\}^n$) to the response space $\mathcal{R}$ (usually $\{0,1\}$).
We call $f$ the true model of a strong PUF since it captures the exact challenge-response behavior. 

ML attacks are usually performed by relying on a functional class of candidate models, collecting CRPs as the training data, and running a learning algorithm to obtain a model from the candidate class which best approximates the true model. 
To claim that a strong PUF is easy to learn, one can propose a learning algorithm which finds a CRP model with good approximation quality using a small number of sample CRPs and terminates in a short time.
The converse is difficult: 
to claim that a PUF is hard to learn, one must show that all possible learning algorithms fail to provide models with good approximation quality, or they require a large number of CRPs or a long running time.

 PAC learning is a known and widely adopted framework for seeking a provable notion of ML resistance with a formal analysis of approximation quality, sample size, and time complexity\footnote{We note that other tools, in addition to PAC theory, for studying ML resistance exist, for example, statistical learning framework \cite{mohri2012foundations}.} \cite{mohri2012foundations}.
We now formalize the passive modeling attack scenario in the context of PAC learning.
A PAC-term for a true model $f$ of a strong PUF is a concept.
Denote as $\mathcal{F}$ the set of all possible PUF-realized functions (every instance of a PUF creates its unique functional mapping $f$).
The set of candidate models used in the learning algorithm is the hypothesis set $\mathcal{H}$. The goal of a learning algorithm is to select a candidate model that matches the true model well. 
Importantly, as shown later, the proof of PAC-hardness guarantees that $\mathcal{H}$ does not have to be restricted to be the same as $\mathcal{F}$ of true models.
This generalization permits a stronger \emph{representation-independent} PAC-hardness proof. While not always possible, representation-independent hardness can be proven for PAC-learning of decryption functions ensuring that no matter how powerful and expressive the chosen $\mathcal{H}$ is, PAC learning decryption function requires exponential time. 

Within the PAC model, CRPs in a training set are assumed to be independent and identically distributed (i.i.d.) under a certain distribution $\mathcal{D}$.

We say a set $\mathcal{F}$ of strong PUFs is PAC-learnable using $\mathcal{H}$, if there exists a polynomial-time algorithm $\mathcal{A}$ such that $\forall \epsilon > 0$, $\forall \delta >0$, for any fixed CRP distribution $\mathcal{D}$, and $\forall f\in\mathcal{F}$, given a training set of size $m$, $\mathcal{A}$ produces a candidate model $h\in\mathcal{H}$ with probability of, at least, $1-\delta$ such that
\begin{equation*}
\Pr_{(\mathbf{c},r)\sim\mathcal{D}}[f(\mathbf{c})\neq h(\mathbf{c})] < \epsilon.
\end{equation*}

In conclusion, our strategy is to say that a strong PUF is ML-resistant if it is not PAC-learnable (i.e., that it is PAC-hard). 
PAC-hardness implies that any successful ML attack requires at least an exponential running time. (We note that PAC theory was used in prior work on PUFs. In \cite{pac_learn_interpose_puf, tamper_proof_sunar, strong_ml_attack_ganji}, the PAC theory was used to show that some PUFs are learnable under the PAC framework. In contrast, we use the PAC framework to establish security guarantees against modeling attacks on a PUF.)

\subsection{Decryption Functions Are not PAC Learnable}

A class of decryption functions of secure public-key cryptosystems has been shown to be not PAC-learnable \cite{kearns1994cryptographic,klivans2006cryptographic}. The proof is outlined below.

A public-key cryptosystem is a triple of probabilistic polynomial-time algorithms $(\Gen,\Enc,\Dec)$ such that:
(1) $\Gen$ takes $n$ as a security parameter and outputs a pair of keys $(pk,sk)$, the public and private keys respectively;
(2) $\Enc$ takes as input the public key $pk$ and encrypts a message (plaintext) $r$ to return a ciphertext $\mathbf{c} = \Enc(pk,r)$;
(3) $\Dec$ takes as input the private key $sk$ and a ciphertext $\mathbf{c}$ to decrypt a message $r = \Dec(sk,\mathbf{c})$.
We only need to discuss public-key cryptosystems encrypting $1$-bit messages ($0$ and $1$).

One of the security requirements of a public-key cryptosystem is that it is computationally infeasible for an adversary, knowing the public key $pk$ and a ciphertext $\mathbf{c}$, to recover the original message, $r$. 
This requirement can also be interpreted as the need for indistinguishability under the chosen plaintext attack (also often referred to as semantic security requirement). 
Given the encryption function $\Enc$ and the public key $pk$, the goal of an attacker is to devise a \emph{distinguisher} $\mathcal{A}$ to distinguish between encryption $\Enc(pk,r)$ of $r=0$ and $r=1$ with non-negligible probability:
\begin{equation*}
\abs{\Pr[\mathcal{A}(pk,\Enc(pk,0))=1] - \Pr[\mathcal{A}(pk,\Enc(pk,1))=1]}\geq \epsilon.
\end{equation*}
A cryptosystem is semantically secure if no polynomial-time attacker can correctly predict the message bit with non-negligible probability.

The connection between the above-stated security of a public-key cryptosystem and the hardness of learning a concept class associated with its decryption function was established in \cite{kearns1994cryptographic,klivans2006cryptographic}. 
The insight of \cite{kearns1994cryptographic,klivans2006cryptographic} is that PAC-learning is a natural result of the ease of encrypting messages with a  public key. 
Since the encryption function $\Enc$ and the public-key $pk$ is known, the distinguishing algorithm can sample independent training examples in the following way: (1) picking a plaintext bit $r$ uniformly randomly from $\{0,1\}$, (2) encrypting $r$ to get the ciphertext $\mathbf{c}=\Enc(pk,r)$. (We later refer to the resulting distribution of ciphertext as the "ciphertext distribution".)
Next, the distinguishing algorithm passes the set of training examples ($(\mathbf{c},r)$'s) into an algorithm for learning the decryption function $\Dec(sk,\cdot)$.
The PAC learning algorithm returns a model $h(\cdot)$ that aims to approximate $\Dec(sk,\cdot)$. 
Using $h(\cdot)$, one could distinguish between ciphertexts stemming from $r=0$ and $r=1$ with non-negligible probability. 
This would entail violating the semantic security of the cryptosystem. 
Technically, this can be summarized as follows \cite{kearns1994cryptographic,klivans2006cryptographic}.
\begin{theorem}
\label{thm:crypto_hardness}
If a public-key cryptosystem is secure against chosen plaintext attacks, then its decryption functions are not PAC-learnable (under the ciphertext input distribution).
\end{theorem}

\subsection{LWE Is Post-Quantum Secure}
\label{sec:lwe_crypto}
According to the cryptographic hardness above, decryption functions of any secure public-key cryptosystem, such as Rivest–Shamir–Adleman (RSA) and elliptic-curve cryptography, can be used to construct ML-resistant PUFs. 
However, integer-factoring-based cryptosystems, including RSA and elliptic-curve cryptography above, become insecure with the development of quantum computers. 
Among all post-quantum schemes, 
the LWE cryptosystem based on hard lattice problems appears to be most promising due to its implementation efficiency and stubborn intractability since 1980s. 

A lattice $\mathcal{L}(\mathbf{V})$ in $n$ dimensions is the set of all integral linear combinations of a given basis $\mathbf{V}=\{\mathbf{v}_1,\mathbf{v}_2,\ldots, \mathbf{v}_n\}$ with $\mathbf{v}_i \in \mathbb{R}^n$:
\begin{equation*}
\mathcal{L}(\mathbf{V}) = \{a_1\mathbf{v}_1 + a_2\mathbf{v}_2+\ldots a_n\mathbf{v}_n: \: \forall a_i \in \mathbb{Z}\}.
\end{equation*}

The LWE problem is defined on the integer lattice $\mathcal{L}(\mathbf{V}) = \{(\mathbf{a},\innerprod{\mathbf{a},\mathbf{s}})\}$ with a basis $\mathbf{V}=(\mathbf{I};\mathbf{s})$, in which $\mathbf{I}$ is an $n$-dimensional identity matrix and $\mathbf{s}$ is a fixed row vector (also called the secret) in $\mathbb{Z}_q^n$.
Throughout this paper, vectors and matrices are denoted with bold symbols with dimension on superscript, which can be dropped for convenience in case of no confusion.
Unless otherwise specified, all arithmetic operations in the following discussion including additions and multiplications are performed in $\mathbb{Z}_q$, i.e. by modulo $q$.

For the lattice $\mathcal{L}(\mathbf{V})= \{(\mathbf{a},\innerprod{\mathbf{a},\mathbf{s}})\}$ with dimension $n$, integer modulus $q$ and a discrete Gaussian distribution $\bar{\Psi}_\alpha$ for noise, the LWE problem is defined as follows.
The secret vector $\mathbf{s}$ is fixed by choosing its coordinates uniformly randomly from $\mathbb{Z}_q$. 
Next $\mathbf{a}_i$'s are generated uniformly from $\mathbb{Z}_q^n$.
Together with the error terms $e_i$, we can compute $b_i = \innerprod{\mathbf{a},\mathbf{s}} + e_i$. 
Distribution of $(\mathbf{a}_i,b_i)$'s over $\mathbb{Z}_q^n\times\mathbb{Z}_q$ is called the LWE distribution $A_{\mathbf{s},\bar{\Psi}_\alpha}$.
The most important property of $A_{\mathbf{s},\bar{\Psi}_\alpha}$ is captured in the following lemma:
\begin{lemma}
\label{lem:LWE_dist}
Based on hardness assumptions of several lattice problems, the LWE distribution $A_{\mathbf{s},\bar{\Psi}_\alpha}$ of $(\mathbf{a},b)$'s is indistinguishable from a uniform distribution in $\mathbb{Z}_q^n\times\mathbb{Z}_q$.
\end{lemma}

Solving the decision version of LWE problem is to distinguish with a non-negligible advantage between samples from $A_{\mathbf{s},\bar{\Psi}_\alpha}$ and those generated uniformly from $\mathbb{Z}_q^n\times\mathbb{Z}_q$. 
This LWE problem is shown to be intractable to solve, without knowing the secret $\mathbf{s}$, based on the worst-case hardness of several lattice problems \cite{regev2009lattices}.
Errors $e$ are generated from a discrete Gaussian distribution $\bar{\Psi}_\alpha$ on $\mathbb{Z}_q$ parameterized by $\alpha >0$: sampling a continuous Gaussian random variable with mean $0$ and standard deviation $\alpha q/\sqrt{2\pi}$ and rounding it to the nearest integer in modulo $q$.  
Notice that error terms are also essential for guaranteeing the indistinguishability: without noise $(\mathbf{a},b)$ becomes deterministic and the secret $\mathbf{s}$ can be solved efficiently via Gaussian elimination methods.

We now describe a public-key cryptosystem based on the LWE problem above in \cite{brakerski2013classical}:
\begin{definition}{(LWE cryptosystem)}
\label{def:LWE_crypto}
\begin{itemize}
\item \textbf{Private key:} $\mathbf{s}$ is uniformly random in $\mathbb{Z}_q^n$ .
\item \textbf{Public key:}  $\mathbf{A}\in \mathbb{Z}_q^{m\times n}$ is uniformly random, and $\mathbf{e}\in \mathbb{Z}_q^m$ with each entry from $\bar{\Psi}_\alpha$. Public key is $(\mathbf{A},\mathbf{b}=\mathbf{A}\mathbf{s}+\mathbf{e})$.
\item \textbf{Encryption:}  $\mathbf{x}\in\{0,1\}^m$ is uniformly random. To encrypt a one-bit plaintext $r$, output ciphertext $\mathbf{c}=(\mathbf{a},b) = (\mathbf{A}^T\mathbf{x},\mathbf{b}^T\mathbf{x}+ r\floor{q/2})$.
\item \textbf{Decryption:} Decrypt the ciphertext $(\mathbf{a},b)$ to $0$ if $b-\innerprod{\mathbf{a},\mathbf{s}}$ is closer to $0$ than to $\floor{q/2}$ modulo $q$, and to $1$ otherwise. 
\end{itemize}
\end{definition}
Notice that each row in the public-key $(\mathbf{A},\mathbf{b})$ is an instance from the LWE distribution $A_{\mathbf{s},\bar{\Psi}_\alpha}$.

Correctness of the LWE cryptosystem can be easily verified: without the error terms, $b-\innerprod{\mathbf{a},\mathbf{s}}$ is either $0$ or $\floor{q/2}$, depending on the encrypted bit. 
Semantic security of the LWE cryptosystem follows directly from the indistinguishability of the LWE distribution from the uniform distribution in $\mathbb{Z}_q^n\times\mathbb{Z}_q$. 
Ciphertexts $(\mathbf{a},b)$ are either linear combinations or shifted linear combination of LWE samples, both of which are indistinguishable from the uniform distribution. 
This is true because shifting by any fixed length preserves the shape of a distribution.
Therefore, an efficient algorithm that can correctly guess the encrypted bit would be able to distinguish LWE samples from uniformly distributed samples. 
This allows \cite{regev2009lattices} to prove that:
\begin{theorem}
\label{thm:LWE_hardness}
Based on the hardness assumptions of several lattice problems, the LWE cryptosystem is secure against the chosen-plaintext attacks using both classical and quantum computers.
\end{theorem}

When the error terms $e_i$'s are introduced:
\begin{align*}
b-\innerprod{\mathbf{a},\mathbf{s}} = &\sum_{i\in S}b_i + \floor{\frac{q}{2}}r - \innerprod{\sum_{i\in S}\mathbf{a}_i,\mathbf{s}}\\
= &\sum_{i\in S}(\innerprod{\mathbf{a}_i,\mathbf{s}}+e_i) - \floor{\frac{q}{2}}r - \innerprod{\sum_{i\in S}\mathbf{a}_i,\mathbf{s}}\\
= &\floor{\frac{q}{2}}r - \sum_{i\in S}e_i,
\end{align*}
in which $S$ is the set of non-zero coordinates in $\mathbf{x}$.
For a decryption error to occur, the accumulated error $\sum_{i\in S}e_i$ must be greater than the decision threshold $\floor{q/4}$. 
The probability of the error is given by \cite{micciancio2009lattice}: 
\begin{align*}
\text{Err}_{\text{LWE}} &\approx 2(1-\Phi(\frac{q/4}{\alpha q \sqrt{m/2}/\sqrt{2\pi}})) \\
&= 2(1-\Phi(\frac{\sqrt{\pi}}{2\alpha\sqrt{m}})),
\end{align*}
in which $\Phi(\cdot)$ is the cumulative distribution function of the standard Gaussian variable. 
We later use this expression to find the practical parameters for the lattice PUF.

\section{Design of Lattice PUF}
\label{sec:design}
\begin{figure}[t!]
    \centering
    \includegraphics[width = 0.8\linewidth]{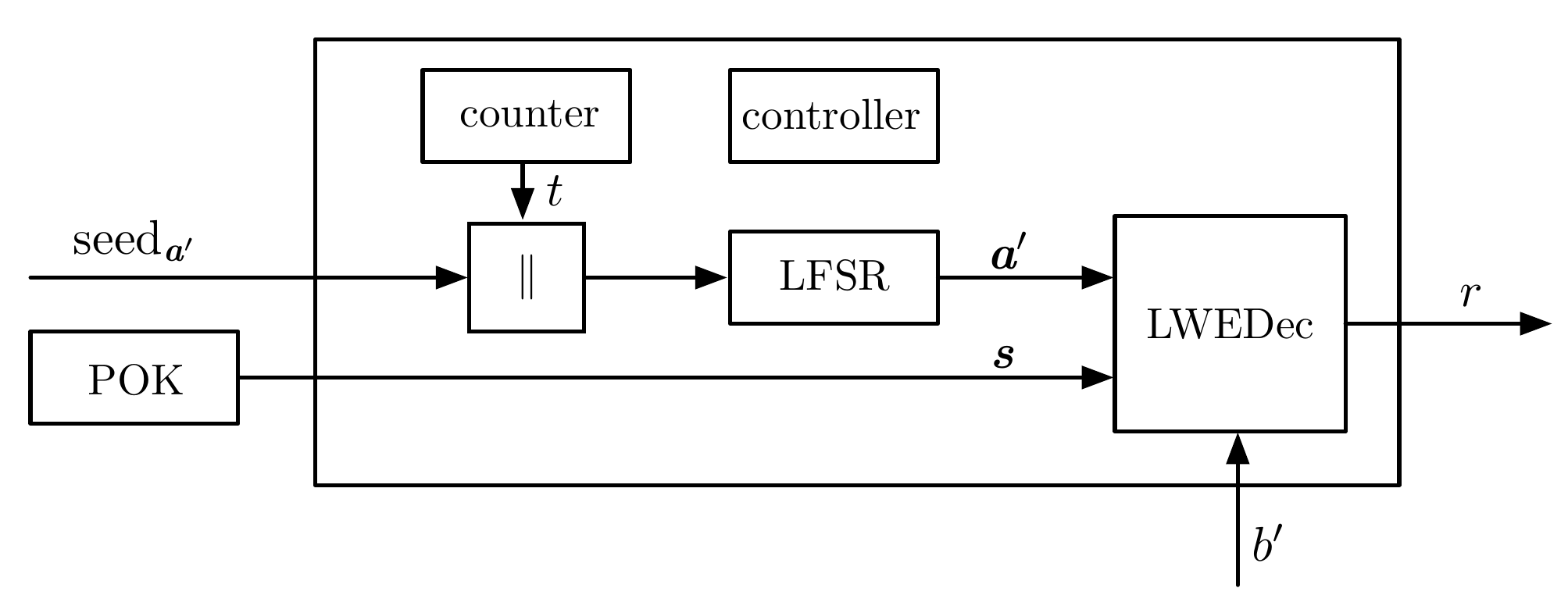}    \caption{Top-level architecture and data flow of the lattice PUF.}
    \label{fig:fpga_impl}
\end{figure}

The top-level architecture of the proposed lattice PUF is shown in Figure \ref{fig:fpga_impl}.

\subsection{Constructing Strong PUF from LWE Decryption Function}
\label{sec:lwe_dec}
\begin{figure}[t!]
    \centering
    \includegraphics[width = 0.8\linewidth]{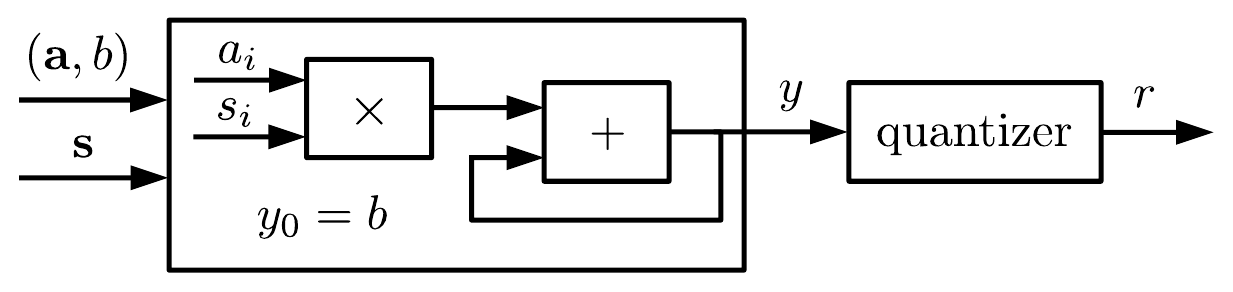}
    \caption{Architecture of LWEDec block.}
    \label{fig:lwedec}
\end{figure}

Figure \ref{fig:lwedec} shows the architecture of LWE decryption (LWEDec) block. 
It takes a binary challenge vector $\mathbf{c} = \{c_0,c_1,\ldots,c_{N-1}\}$ of size $N = (n+1)\log q$ which maps to a ciphertext $(\mathbf{a},b)$ in the following way:
\begin{align*}
a_i &= \sum_{j=0}^{\log q-1}c_{(i-1)\log q+j}2^j,\; \forall i\in \{1,2,\ldots,n\}, \\
b &= \sum_{j=0}^{\log q-1}c_{n\log q+j}2^j. 
\end{align*}
Here $a_i$ denotes the $i$-th element of the integer vector $\mathbf{a}\in\mathbb{Z}_q^n$.
In this paper, without specification, $\log(x)$ refers to $\log_2(x)$.
Similarly, the private key $\mathbf{s}$ for the corresponding LWE decryption function is realized by a binary secret key $\mathbf{W} =\{W_0,W_1,\ldots,W_{n\log q-1}\} $ of size $n\log q$:
\begin{equation*}
s_i=\sum_{j=0}^{\log q-1} W_{(i-1)\log q+j} 2^j,\; \forall i\in \{1,2,\ldots,n\}.
\end{equation*}
A modulo-dot-product $b-\innerprod{\mathbf{a},\mathbf{s}}$ is computed using the modulo-MAC unit. 
It can be implemented in a serial way using $n$ stages. 
Recall that all additions and multiplications are performed in modulo $q$.
Since $q$ is a power of $2$ in our construction, modulo addition and multiplication can be naturally implemented by integer addition and multiplication that keep only the last $\log q$-bit result. 
Finally the response $r$ is produced by a quantization operation $r = Q(b-\innerprod{\mathbf{a},\mathbf{s}})$: 
\begin{equation*}
Q(x) = \begin{cases}
	0& x \in [0,\frac{q}{4}]\cup(\frac{3q}{4},q-1],\\
	1& x \in (\frac{q}{4},\frac{3q}{4}].
\end{cases}
\end{equation*}

The computation above can be directly implemented as a strong PUF with $2^N$ CRPs since it maps a challenge vector $\mathbf{c}\in \{0,1\}^N$ into a binary response $r\in\{0,1\}$.
We now discuss parameter selection for the LWE decryption function. 
In general, we seek to find design parameters such that 
(1) the resulting PUF has excellent statistical properties, such as uniformity, uniqueness, and reliability, 
(2) successful ML attacks against it require an un-affordably high time complexity in practice, and 
(3) its hardware implementation costs are minimized. 

Prior theoretical arguments establish the impossibility of a polynomial-time attacker. 
To guarantee practical security, we need to estimate the number of samples and the actual running time (or a number of CPU operations) required for a successful ML attack. \cite{regev2009lattices} shows that a small number of samples are enough to solve an LWE problem, but in an exponential time. 
Thus, we refer to runtime as concrete ML resistance (or ML hardness) and say that a PUF has $k$-bit ML resistance if any successful ML attack requires at least $2^k$ operations. 
We adopt the estimator developed by Albrecht \emph{et al.} \cite{albrecht2015concrete} to estimate concrete ML hardness. 
The concrete hardness of an LWE problem increases with the increase of LWE parameters $n$, $q$, and $\alpha$ for all types of attacks.
Recall that $n$ represents the lattice dimension, $q$ represents the range of integer for each dimension, and $\alpha$ reflects the noise level in CRP (ciphertext) generation.
For a given set of parameters, the estimator compares the complexity of several most effective attacks, including decoding, basis reduction, and meet-in-the-middle attacks  
\cite{howgrave2007hybrid,lindner2011better}. 
We utilize the estimator in a black-box fashion to find the set of parameters with the target of $128$-bit concrete ML resistance. 

We consider two metrics of implementation cost, both of which scale with $n$: the number of challenge and secret bits needed ($n\log q$), and the number of MAC operations ($n$).
This motivates the need to decrease $n$.

\begin{figure}[t!] 
    \centering
  \subfloat[\label{fig:1_bit_hw}]{%
       \includegraphics[width=0.48\linewidth]{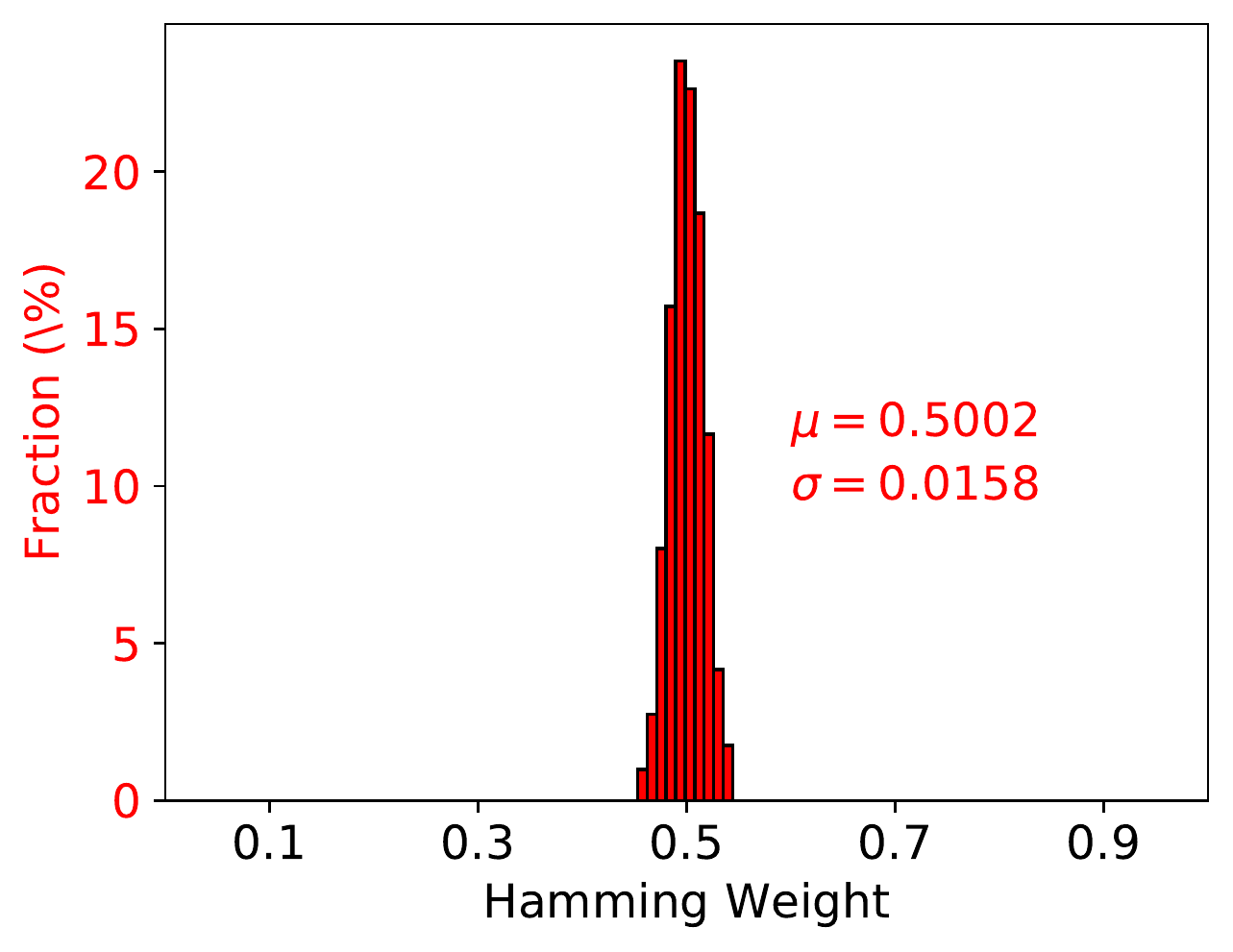}}
    \hfill
  \subfloat[\label{fig:num_POK_vs_dec_err}]{%
        \includegraphics[width=0.5\linewidth]{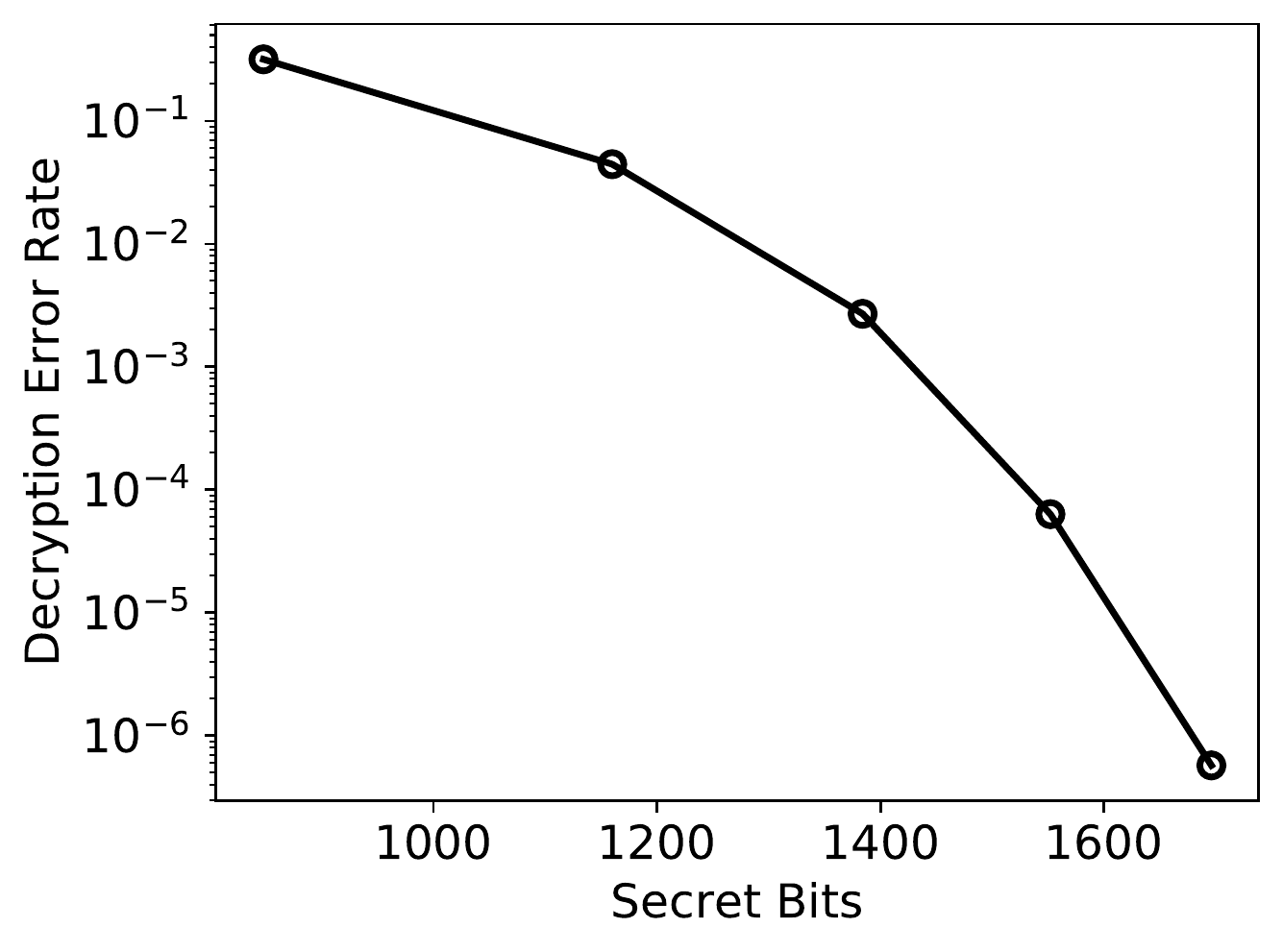}}
  \caption{(a) CRPs change radically even with 1 bit POK change. (b) Super-exponential decrease of decryption error rate with the increase of secret bits. The analysis is done for 128-bit concrete hardness.}
  \label{fig: lattice_puf_features} 
\end{figure}

Output errors of the lattice PUF come from two sources: (1) environmental errors of secret bits, and (2) errors of decryption during response generation.
The former can be thought as the failure of key reconstruction in POKs.
Figure \ref{fig:1_bit_hw} shows the hamming-distance between CRPs generated using error-free POKs and POKs with 1 bit error.
Since a single bit-flip completely changes the challenge-response behavior of LWE decryption function, the failure rate of key reconstruction needs to be low, e.g. $10^{-6}$ (as widely adopted in other PUF applications \cite{maes2012pufky}).
Section \ref{sec:result} describes how the target failure rate can be achieved via a conventional FE based on the error-correcting codes.
The latter corresponds to the decryption error and is orthogonal to errors in the secret key $\mathbf{s}$. 
Recall that in CRP generation of the lattice PUF, a bit of plaintext $r$ is sampled and the ciphertext $\mathbf{c}$ is produced by a noisy encryption function $\mathbf{c}=\Enc(r)$. 
Given ciphertext $\mathbf{c}$ as input challenge, the decryption function can output a wrong response $\tilde{r}\neq r$ when the accumulated error $\sum_{i\in S} e_i$ in the encryption function exceeds the decision boundary.

In order for a strong PUF to be used in direct authentication, its decryption error rate should be small enough for reliable distinguishability of long strings.
We set the target around $2\%$.
Figure \ref{fig:num_POK_vs_dec_err} explores the trade-off between the number of secret bits and the decryption error rate needed for $128$-bit concrete ML hardness.
It shows that, at fixed concrete ML hardness, the decryption error rate decreases super exponentially with the number of secret bits. 

Considering the design metrics above, a feasible set of parameters is found using the estimator in \cite{albrecht2015concrete}.
By setting $n=160$, $q=256$, $m=256$ and $\alpha = 2.20\%$, we achieve a lattice PUF with $128$-bit concrete hardness and a decryption error rate of $1.26\%$.

In order to get a $1$-bit response, $N = (n+1)\log q = 1288$ bits need to be sent to the lattice PUF as a challenge.
For direct authentication applications, usually around $L = 100$ bits of responses are required. Therefore, the direct implementation described so far would require $128.8K$ challenge bits.
This high ratio of challenge length to response length limits its practical use in many scenarios when communication is expensive.

\vspace{-0.5em}
\subsection{Challenge Compression through Distributional Relaxation}
\label{sec:lfsr}

The LWEDec block described in Section \ref{sec:lwe_dec} requires a challenge $\mathbf{c}$ in the form $\mathbf{c}=(\mathbf{a},b)$ to be sent from the server to the PUF. 
To represent vector $\mathbf{a} \in \mathbb{Z}_q^n$ requires $n\log q$ bits while to represent scalar $b \in \mathbb{Z}_q$  requires only $\log q$ bits. 
Thus, the major cost of transmission lies in sending $\mathbf{a}$. 
We wish to avoid sending $\mathbf{a}$ directly and, instead, to send a compressed (shorter) version of $\mathbf{a}$  and re-generate its full-size version on the PUF.
Our approach is enabled by the recent results on the distributional behavior of $\mathbf{a}=\mathbf{A}^T\mathbf{x}$ \cite{akavia2009simultaneous} and the concept of space-efficient LWE \cite{galbraith2013space}.

Recall that $b$ is given by: 
\begin{align*}
    b&=\mathbf{b}^T\mathbf{x}+r\floor{q/2}\\
    &= (\mathbf{A}\mathbf{s}+\mathbf{e})^T\mathbf{x}+r\floor{q/2}\\
    &=(\mathbf{A}^T\mathbf{x})^T\mathbf{s}+\mathbf{e}^T\mathbf{x}+r\floor{q/2}.
\end{align*}
First, we replace the component $\mathbf{a}=\mathbf{A}^T\mathbf{x}$ by $\mathbf{a}^*$ uniformly randomly sampled from $\mathbf{Z}_q^n$. That allows us to represent challenge $\mathbf{c} = (\mathbf{a},b)$:
\begin{equation*}
    \begin{cases}
    \mathbf{a}= \mathbf{A}^T\mathbf{x}\\
    b = (\mathbf{A}^T\mathbf{x})^T\mathbf{s}+\mathbf{e}^T\mathbf{x}+r\floor{q/2}
    \end{cases}
\end{equation*}
as $\mathbf{c}^*=(\mathbf{a}^*,b^*)$:
\begin{equation*}
    \begin{cases}
    \mathbf{a}^*\\
    b^*=\mathbf{a}^{*T}\mathbf{s}+\mathbf{e}^T\mathbf{x}+r\floor{q/2}
    \end{cases}.
\end{equation*}
In \cite{akavia2009simultaneous}, it is proven that distribution of $\mathbf{c}^*=(\mathbf{a}^*,b^*)$ is statistically close to the original ciphertext distribution, therefore the required security properties are preserved.  

The advantage of the above approximation is that, as shown by \cite{galbraith2013space}, several low-complexity  PRNGs are capable of producing an output string $\mathbf{a}^\prime$ suitably close to $\mathbf{a^*}\in\mathbb{Z}^n_q$ within the context of LWE cryptosystem. In particular, an LFSR is an especially simple PRNG having the right properties.
Specifically, a vector $\mathbf{a}^\prime$ generated by an LFSR provides similar concrete security guarantees against standard attacks on LWE, such as CVP reduction, decoding, and basis reduction \cite{galbraith2013space}.
This is because LFSR-generated $\mathbf{a}^\prime$ maintains good properties including:
\begin{itemize}
    \item it is hard to find ``nice'' bases for a lattice with basis from LFSR-generated $\mathbf{a}^\prime$;
    \item given an arbitrary vector in $\mathbb{Z}_q^n$, it is hard to represent it as a binary linear combination of LFSR-generated $\mathbf{a}^\prime$'s;
    \item it is hard to find a short vector $\mathbf{w}$ that is orthogonal to LFSR-generated $\mathbf{a}^\prime$'s.
\end{itemize}

The ability to rely on a simple PRNG to produce $\mathbf{a}^\prime$ allows a dramatic reduction in challenge transfer cost. 
Now, the challenge $\mathbf{c}^\prime$ contains only a small $\text{seed}_{\mathbf{a}^\prime}$ into the PRNG and the corresponding $b^\prime$ as
\begin{align*}
    b^\prime&=(\mathbf{a}^\prime)^T\mathbf{s}+\mathbf{e}^T\mathbf{x}+r\floor{q/2}\\
    &= \LFSR{(\text{seed}_{\mathbf{a}^\prime})}^T\mathbf{s}+\mathbf{e}^T\mathbf{x}+r\floor{q/2}.
\end{align*}
Here $\LFSR(\cdot)$ denotes the output generated by an LFSR.

With LWE parameters chosen as Section \ref{sec:lwe_dec}, using a seed of length $256$ is able to reduce the challenge length from $1288$ to $256+8=264$ per one bit of response. 
The improvement of efficiency becomes more pronounced for generating multi-bit responses:
This is because $\mathbf{a}^\prime_1, \mathbf{a}^\prime_2 \ldots \mathbf{a}^\prime_t$ can be generated sequentially from a shared seed, so that only the seed and $b^\prime_1, b^\prime_2, \ldots, b^\prime_t \in Z_q$ are required to be sent to the PUF side. 
$100$ bits of responses now require only transmitting $256+100\times\log 256 = 1056$ bits for challenges.

\subsection{Countermeasure for Active Attack}
\label{sec:counter}

We now demonstrate an active attack that compromises security of lattice PUF. The attack is premised on the ability to supply arbitrary challenges (ciphertexts) as inputs to the decryption function. 
The attack proceeds as follows. 
The attacker fixes $\mathbf{a}$ and enumerates all possible $b\in \mathbb{Z}_q$ for challenge $\mathbf{c} = (\mathbf{a},b)$.
As $b$ increases from $0$ to $q-1$, the response $r = Q(b-\innerprod{\mathbf{a},\mathbf{s}})$ changes from  $Q(b-\innerprod{\mathbf{a},\mathbf{s}}) = 0$ to $Q(b+1-\innerprod{\mathbf{a},\mathbf{s}}) = 1$ exactly when $b$ satisfies
\begin{equation*}
b-\innerprod{\mathbf{a},\mathbf{s}} = q/4.
\end{equation*}
We denote this specific value of $b$ as $\hat{b}$. 
The exact value of $\innerprod{\mathbf{a},\mathbf{s}}$ can then be extracted by $\innerprod{\mathbf{a},\mathbf{s}} = \hat{b} - q/4$. 
By repeating this procedure $n$ times, the attacker is able to set up $n$ linear equations (without errors):  
\begin{align*}
\label{eq:secret_eqs}
    \innerprod{\mathbf{a}_0,\mathbf{s}} &= \hat{b}_0 - q/4, \\
    \innerprod{\mathbf{a}_1,\mathbf{s}} &= \hat{b}_1 - q/4, \\
    & \cdots \\
    \innerprod{\mathbf{a}_{n-1},\mathbf{s}} &= \hat{b}_{n-1} - q/4.
\end{align*}
Gaussian elimination can then be used to solve for $\mathbf{s}$, entailing compromising the system. 

To defend against the attack, we introduce a self-incrementing counter to embed the counter value into a challenge seed \cite{yu2016lockdown}.
This makes the attack impossible as the counter restricts the attacker's ability to completely control input challenges to the LWEDec block.
As a result, the attacker cannot enumerate all values of $b$ while keeping $\mathbf{a}$ unchanged. 
As shown in Figure \ref{fig:fpga_impl}, the concatenation of the challenger-provided seed and the counter value $t$ (i.e. $\text{seed}_{\mathbf{a}'}||t$) is used as the seed for generating $\mathbf{a}$. 
The counter value is public and is incremented by $1$ on each response generation.

\subsection{Latency Optimization via Design Parallelization}
\label{sec: lpuf_par}

The lattice PUF architecture shown in Fig. \ref{fig:fpga_impl} achieves low hardware implementation cost via a bit-serial design. 
However, the highly serialized design leads to inefficient utilization of clock cycles resulting in large response latency. To generate one bit of response, the design needs $160$ sequential MAC operations and each MAC needs to wait $8$ cycles until one byte of ciphertext is generated by the bit-serial LFSR. This is undesirable in performance-critical applications.

We observe that latency is limited by two factors: (1) the LFSR and LWEDec datapath (LFSR-LWEDec) produces response bits serially, and (2) the LWEDec block performs a single MAC sequentially.  
We optimize each aspect.

\begin{figure}[t!]
\centering
\includegraphics[width = 0.8\linewidth]{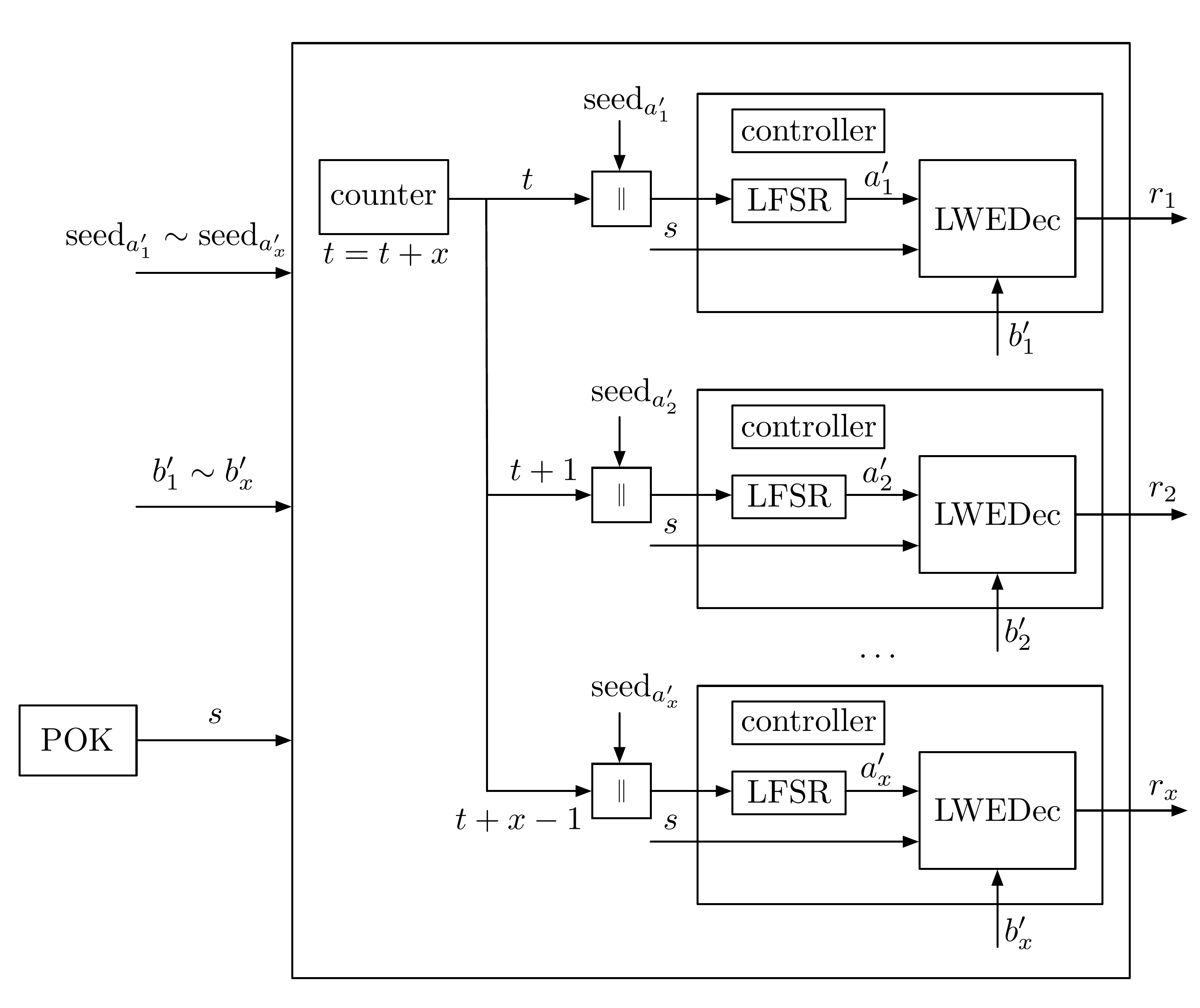}
\caption{Reducing latency via a parallel LFSR-LWEDec datapath.}
\label{fig:lpuf_p1}
\end{figure}

We first parallelize the LFSR-LWEDec data-path which implements the functionality in response generation: the LFSR generates ciphertext and LWEDec block performs modulo MAC operation. 
Parallelizing it allows multiple response bits to be generated simultaneously, Fig. \ref{fig:lpuf_p1}. Each LFSR-LWEDec data-path uses the same POK, but receives different LFSR seeds and ciphertext inputs. 

We adapt the counter technique described in Section \ref{sec:counter}.  
We embed an increasing counter value into the seed of each datapath, Fig. \ref{fig:lpuf_p1}. 
The strategy directly enables $P_1$ times reduction in response generation latency, at the cost of $P_1$ times increase in seed transmission and datapath hardware utilization.

We now investigate how to increase the dot-product throughput in LWEDec. 
Since the LWEDec block contains only a single MAC unit, we increase the number of MACs. However, the throughput of the LFSR limits the overall throughput of the data-path: it produces only a single output bit per cycle. 
Therefore, to fully utilize the parallel MAC units, the LFSR needs to produce a sufficient number of ciphertext bytes on each cycle. We adopt an unrolled LFSR to increase its throughput.

\begin{figure}[t!]
\centering
\includegraphics[width = 0.8\linewidth]{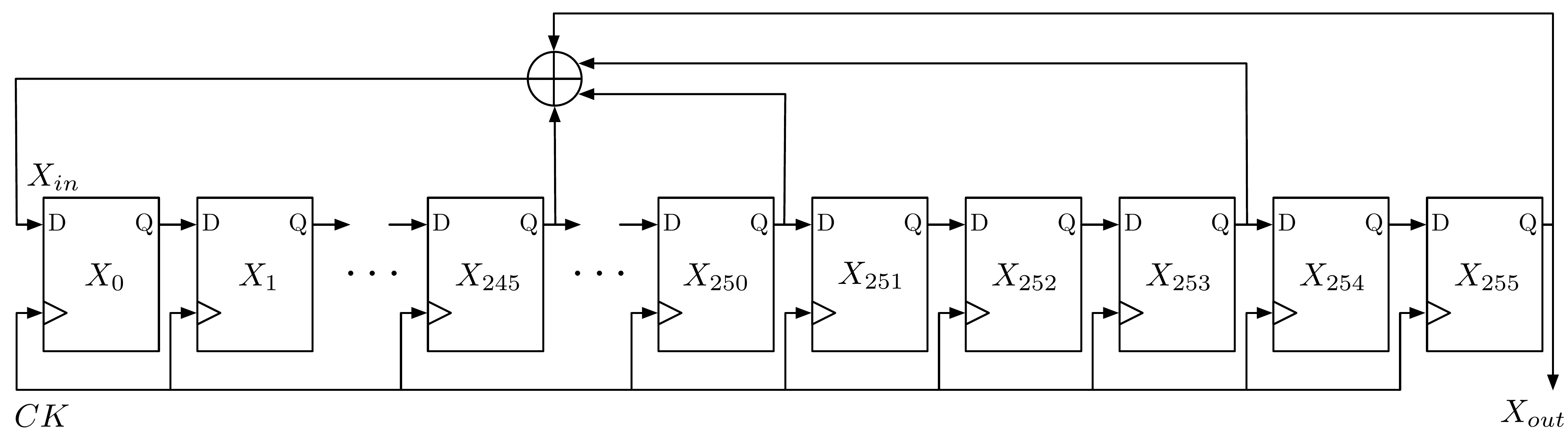}
\caption{A 256-bit bit-serial LFSR.}
\label{fig:lfsr_baseline}
\end{figure}

\begin{figure}[t!]
\centering
\includegraphics[width = 0.8\linewidth]{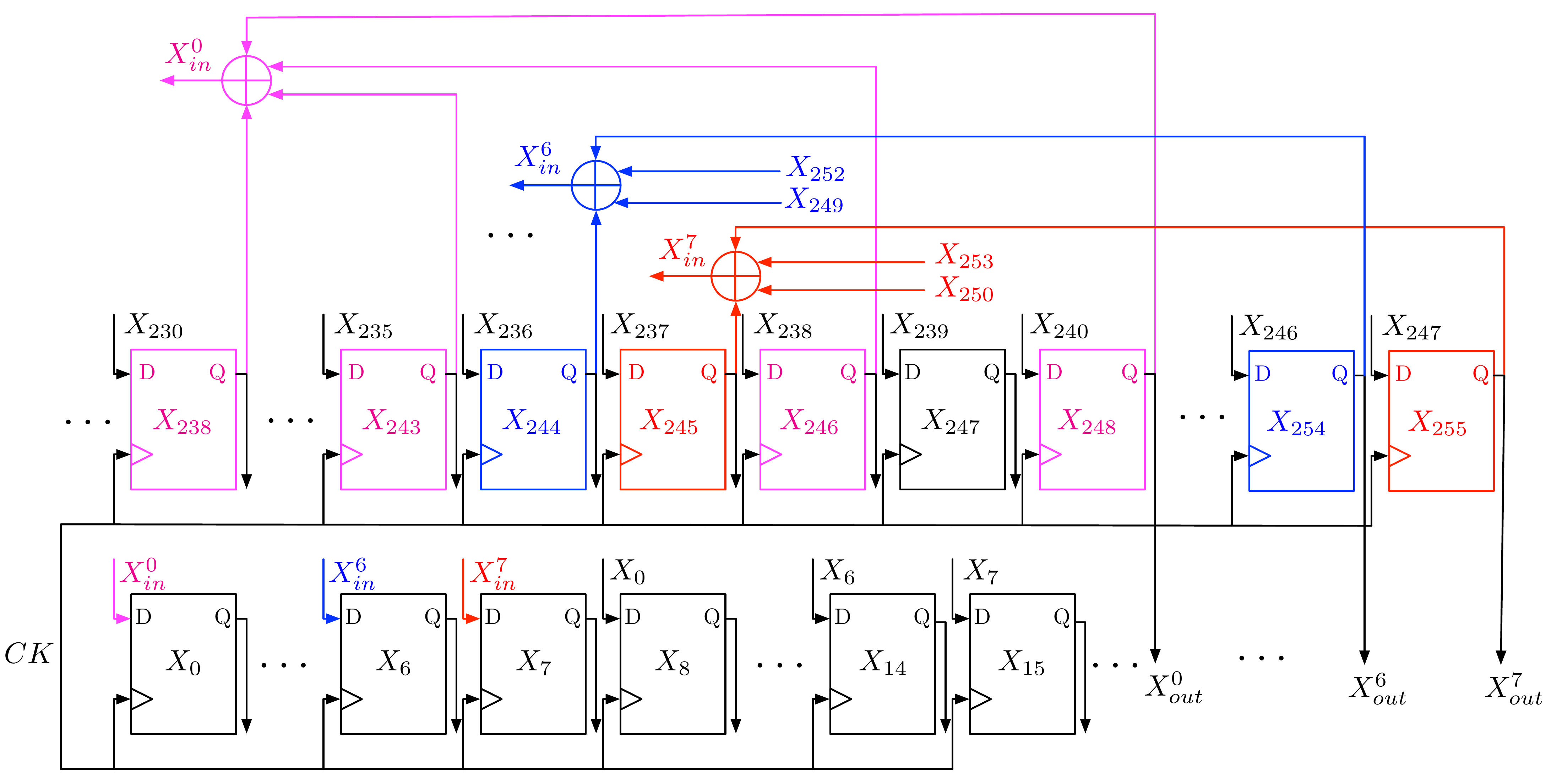}
\caption{A 256-bit LFSR unrolled by a factor of 8.}
\label{fig:lfsr_unrolled}
\end{figure}

\begin{figure}[t!]
\centering
\includegraphics[width = 0.8\linewidth]{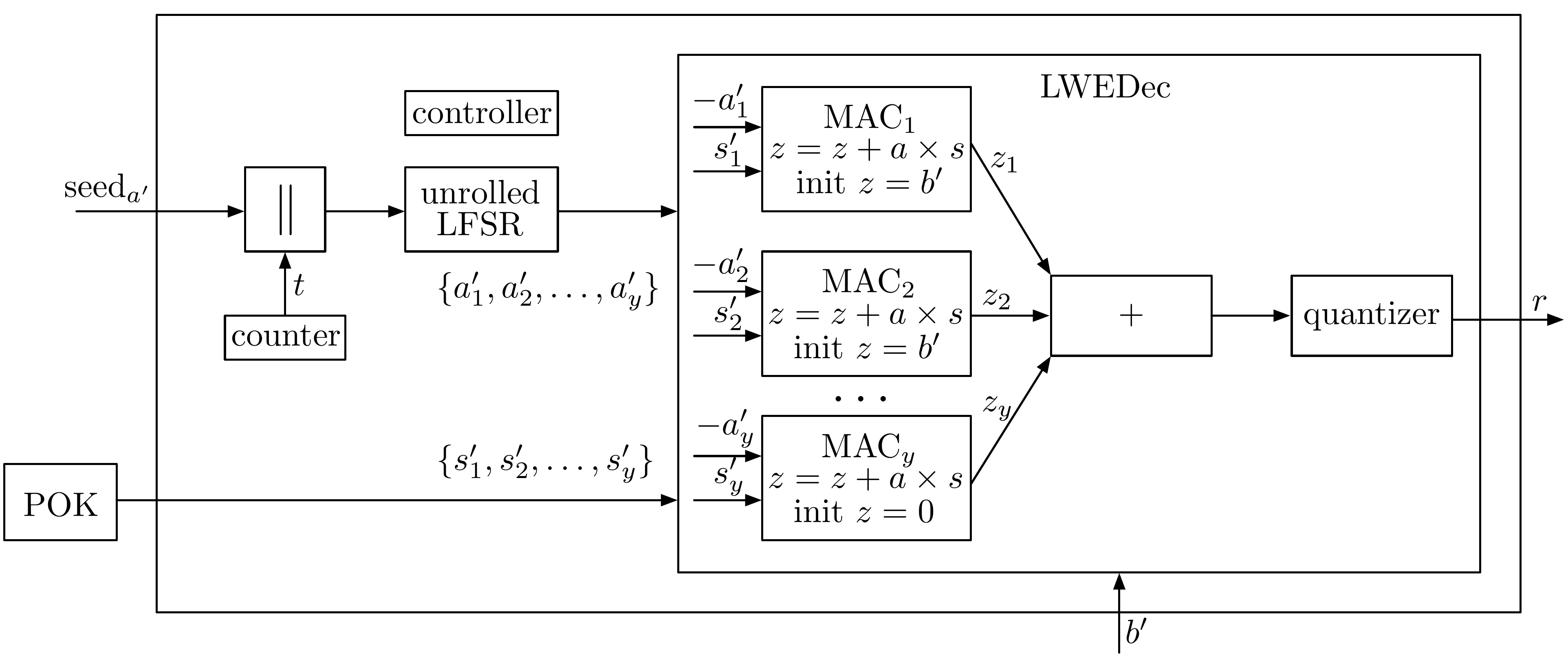}
\caption{Reducing latency via parallel MAC units.}
\label{fig:lpuf_p2}
\end{figure}

An unrolled LFSR is functionally equivalent to the bit-serial LFSR, capable of producing multiple bits per cycle 
\cite{cycle_efficient_lfsr, unrolled_lfsr_ref_2}. We adopt the basic loop unrolling strategy \cite{cycle_efficient_lfsr} which completes the compute of multiple cycles within one cycle.

On each cycle, a bit-serial LFSR typically shifts the register bits by one position and shifts in one feedback bit, which is the XOR of higher register bits. A single output bit is produced. The basic idea of unrolling a LFSR is to parallelize the compute of multiple cycles to produce multiple output bits in one cycle. On each cycle, an unrolled LFSR completes the following tasks: computing the feedback bits of multiple consecutive cycles, shifting the register bits by multiple positions, loading the feedback bits to the multiple lowest registers, and assigning multiple output bits. 

We demonstrate how to unroll the 256-bit LFSR by a factor of 8. Fig. \ref{fig:lfsr_baseline} shows the LFSR implementation. The feedback bit is defined as $X_{in}$ and the output bit is defined as $X_{out}$. 
We first compute feedback bits for 8 consecutive cycles:
\begin{align*}
    X_{in} ^ 7 &=  X_{255} \oplus X_{253} \oplus X_{250} \oplus X_{245} \\
    X_{in} ^ 6 &=  X_{254} \oplus X_{252} \oplus X_{249} \oplus X_{244} \\
    &\cdots \\
    X_{in} ^ 0 &= X_{248} \oplus X_{246} \oplus X_{243} \oplus X_{238}
\end{align*}

Then we modify the stride of the shift to update the register state after multiple single-position shifts. Besides the lowest 8 register bits, each register bit is updated with the bit value, located 8 bits away from the current bit:

\begin{equation*}
    X_{i} = X_{i-8}
\end{equation*}

The lowest 8 register bits latch values of the 8 feedback bits $X_{in} ^ 7$ to $X_{in} ^ 0$. Finally, we assign multiple register bits to output bits:
\begin{align*}
    X_{out} ^ 7 &=  X_{255} \\
    X_{out} ^ 6 &=  X_{254}  \\
    &\cdots \\
    X_{out} ^ 0 &= X_{248} 
\end{align*}

\begin{figure}[t!]
\centering
\includegraphics[width = 0.8\linewidth]{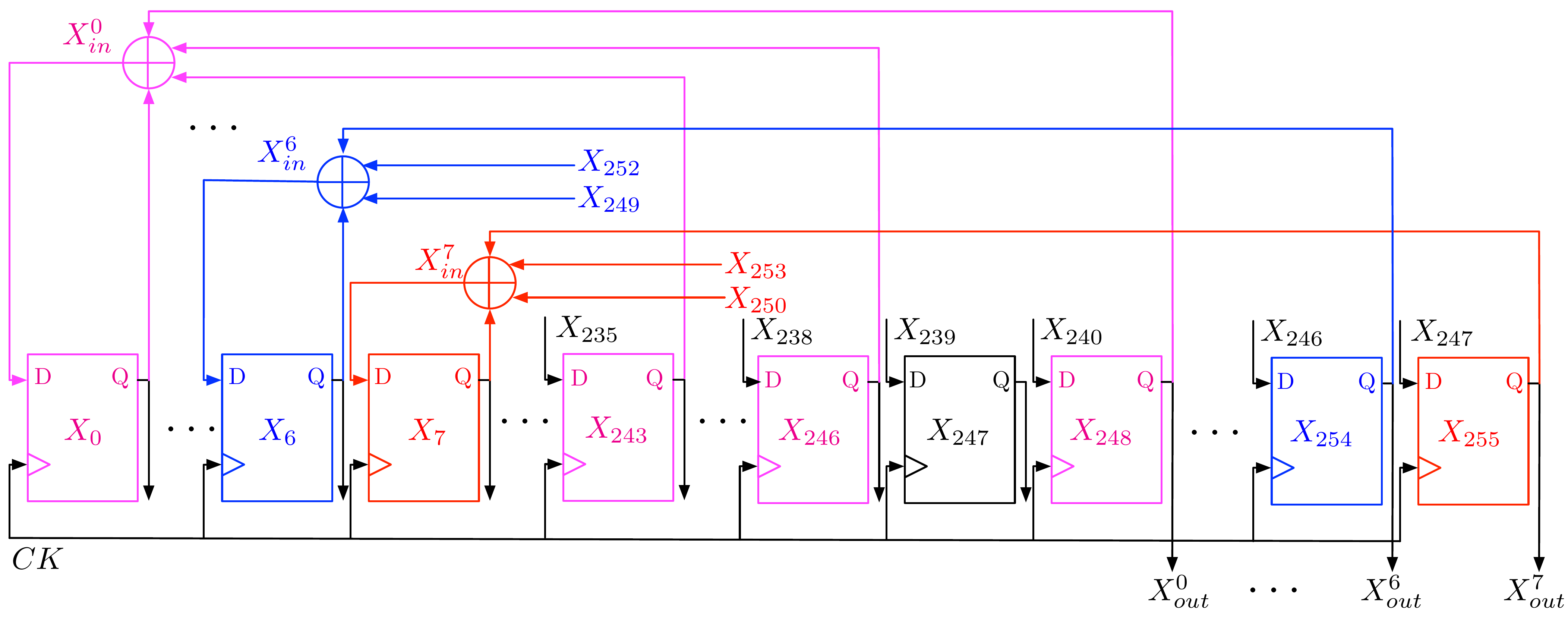}
\caption{The LFSR generator polynomial limits the maximum unrolling factor. It cannot produce feedback bits for more than $8$ consecutive cycles due to the constraint of $X_{7}$.}
\label{fig:lfsr_unrol_limit}
\end{figure}

Fig. \ref{fig:lfsr_unrolled} shows the architecture of the unrolled LFSR. It improves the throughput of the baseline bit-serial LFSR by 8X, at the cost of more XOR gates for computation.

We implemented parallel MAC units in the LWEDec block with an unrolled LFSR generating multiple bytes of ciphertext, Fig. \ref{fig:lpuf_p2}. 
We initialized the accumulation register of one of the MAC unit with ciphertext $b$, and the accumulation registers of the rest of MAC units to 0. 
On each cycle, each MAC unit accumulates the product of a byte of POK with the opposite of one byte of ciphertext $\mathbf{a}$ 
(Recall that the LWE decryption function computes $b-\innerprod{\mathbf{a},\mathbf{s}}$). 
Finally the partial sums from each MAC unit are accumulated together to produce a dot-product result. The result is quantized and a response bit is produced. The increased number of MAC units reduces the number of MAC operations assigned to each unit, thus reducing the dot-product latency. On a design with $P_2$ parallel MAC units, the strategy reduces the dot-product latency for a response bit by a factor of $P_2$. The cost is the increase in hardware resource utilization due to an unrolled LFSR and the parallel MAC units.

The LFSR-LWEDec datapath parallelization and MAC unit parallelization can be combined. 
Specifically, each parallel LFSR-LWEDec datapath can adopt an unrolled LFSR and a LWEDec block with multiple MAC units. 
In practice, a user selects $P_1$ and $P_2$ values based on the latency requirement, resource budget, and the baseline LFSR implementation. MAC unit parallelization is more resource-efficient since the strategy only needs additional hardware for unrolled LFSR and MAC units. In contrast, the LFSR-LWEDec data-path parallelization requires duplication of the entire datapath. 

However, the MAC unit parallelization cannot achieve arbitrary parallelism since the LFSR cannot be unrolled with an arbitrary factor. The LFSR generator polynomial limits the maximum unrolling factor, Fig. \ref{fig:lfsr_unrol_limit}. If the 256-bit LFSR takes the XOR result of $X_{255}$, $X_{253}$, $X_{250}$ and $X_{7}$ as the feedback bit, the LFSR cannot be unrolled by over 8 times with the demonstrated technique, since the LSB $X_{0}$ is only $7$ bits away from $X_{7}$. This indicates the unrolled LFSR cannot produce more than $8$ consecutive feedback bits utilizing consecutive register bits lower than $X_{7}$. The lowest LFSR bit involved in the feedback bit calculation determines the maximum unrolling times. LFSR-LWEDec data-path parallelization is a generic strategy. Its maximum parallelism is not constrained by any specific LFSR implementation. Therefore, in practice, the optimal strategy is to first seek parallelization via LFSR unrolling, and then parallelize the LFSR-LWEDec data-path to achieve further latency optimization once the unrolling of LFSR reaches the limit. We validate the strategy with a detailed design space exploration in Section \ref{sec:result}.

\begin{figure}[t!]
\centering
\includegraphics[width = 0.8\linewidth]{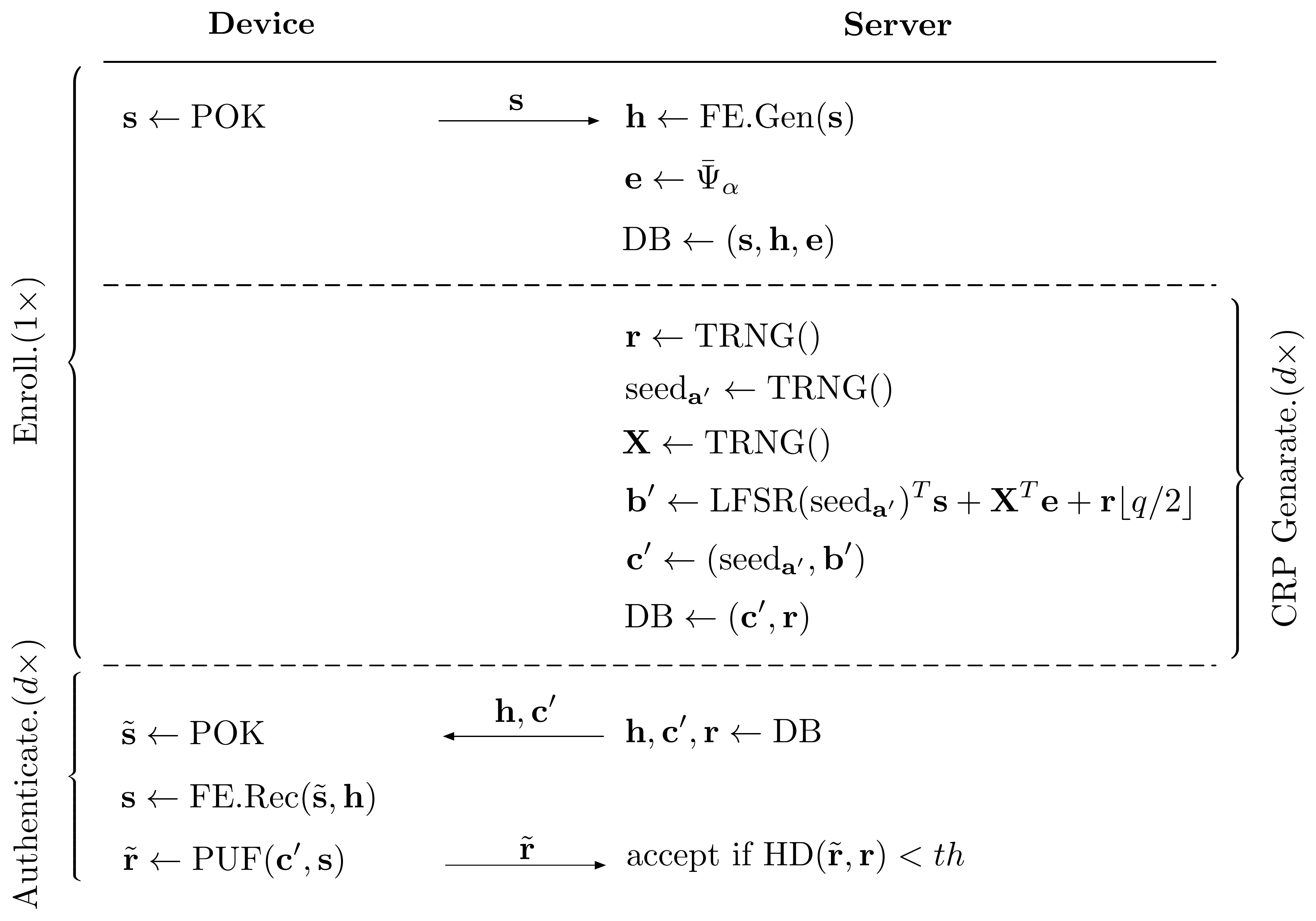}
\caption[]{End-to-end authentication method with the lattice PUF.\footnotemark}
\label{fig:protocol}
\end{figure}

\subsection{PUF-Based Authentication}
As an example application, we study using the lattice PUF in a string-matching authentication scheme \cite{suh2007physical}. 
Figure \ref{fig:protocol} describes an end-to-end authentication method using the lattice PUF. 

A fuzzy extractor \footnotetext{We note that the scheme demonstrated here is slightly different from \cite{lpuf_tc_published}. This scheme adopts the more conventional fuzzy extractor technique for POK reconstruction, in contrast to the reverse fuzzy extractor technique in \cite{lpuf_tc_published}.} \cite{bosch2008efficient} is used for reconstructing a stable secret key from the POK bits.
During the enrollment phase, the POK bits, based on the power-up values of SRAM cells, are read out through a one-time interface and sent to the server. The fuzzy extractor generates helper data (\texttt{FE.Gen})
for subsequent reconstruction, and both the secret key $\mathbf{s}$ and helper data $\mathbf{h}$ are stored in the server database (\texttt{DB}).

Note that there is a critical difference between the PUF and the public key cryptosystem use-cases. Conventional applications using a public key cryptosystem assume that only an insecure public channel is available to communicating parties. 
This requires that public key generation takes place on the device. In a LWE-based PKE, that requires implementing on the device the relatively expensive discrete Gaussian sampling mechanism. In contrast, a typical assumption of secure protocols utilizing PUF-based authentication is that during the \emph{enrollment phase}, secret information can be transferred in a secure manner. 
With $\mathbf{s}$ securely transferred to the server side during the enrollment phase, the computationally costly discrete Gaussian sampling ($\mathbf{e}$) can happen on the server rather than the PUF (device) side.
We highlight again that it is the availability of $\mathbf{s}$ on the server that enables challenge compression via distributional relaxation and the corresponding design improvements. 

Each authentication transaction is based on getting a string of response bits of length $L$ from a PUF. 
In the CRP generation phase, we use the challenge compression technique described in Section \ref{sec:lfsr}, to generate a single challenge $\mathbf{c}^\prime$ that produces an $L$-bit response string. 
For clarity, we refer to such a challenge as an $L$-bit (output) CRP in contrast to a basic single-bit (output) CRP discussed earlier. 
The $L$-bit CRP generation starts from sampling a string $\mathbf{r}$ of length $L$, a starting seed $\text{seed}_{\mathbf{a}^\prime}$, and a binary matrix $\mathbf{X}$ of dimension $m\times L$ via a true random number generator (TRNG). 
Next, $\text{seed}_{\mathbf{a}^\prime}$ is expanded by LFSR to a matrix of dimension $n\times L$ and a vector $\mathbf{b}^\prime$ of length $L$ is calculated accordingly. 
The challenge of the $L$-bit CRP is $\mathbf{c}^\prime = (\text{seed}_{\mathbf{a}^\prime},\mathbf{b}^\prime)$.
Finally, this L-bit CRP $(\mathbf{c}^\prime, \mathbf{r})$ is stored in DB. 

During each authentication transaction, the server sends a valid challenge $\mathbf{c}^\prime=(\text{seed}_{\mathbf{a}^\prime}, \mathbf{b}')$ of an $L$-bit CRP together with a device-specific helper data to the device. 
To avoid device impersonation via replay, the CRPs are used only once during authentication. Therefore, the number of stored CRPs and thus possible authentications $d$ should be sufficiently large.
If the HD between $\mathbf{s}$ and $\tilde{\mathbf{s}}$ is within the error-decoding capability of the error correction code (ECC) decoder, the decoder reconstructs the correct secret key $\mathbf{s}$ (\texttt{FE.Rec}). 
The device then generates the PUF response string $\tilde{\mathbf{r}}$ by executing the LWE decryption function with $(\mathbf{c}^\prime, \mathbf{s})$ and sends it back to the server. 
The server compares the $L$-bit $\tilde{\mathbf{r}}$ string with the corresponding $L$-bit $\mathbf{r}$ string stored in DB:
the device is accepted if $\text{HD}(\tilde{\mathbf{r}}, \mathbf{r})$ is less than a given threshold $th$.

\section{Experimental Results}
\label{sec:result}

We simulate the behavior of the proposed ML-resistant PUF using Python. 
The statistics of raw SRAM PUF responses are based on \cite{maes2013accurate, maes2009soft}. 
The logical behavior of other digital circuits are accurately captured by software. 
We virtually manufacture (simulate) $1000$ distinct lattice PUF instances with design parameters chosen in Section \ref{sec:design}. 
The CRPs of the PUF instances are collected to examine: (1) statistical characteristics (uniformity, uniqueness, and reliability), and (2) resistance to leading ML modeling attacks. 
The whole lattice PUF design (besides the SRAM PUF) is implemented on a Xilinx Spartan-6 FPGA.

\subsection{Statistical Analysis}
\label{sec:statistical_result}

In this section, we examine statistical metrics of the proposed PUF. 
Uniformity, uniqueness and reliability are commonly adopted metrics to evaluate the statistical performance of a PUF \cite{maiti2013systematic}. 
\textbf{Uniformity} measures the mean Hamming Weight of responses from a PUF under a set of random challenges. 
The ideal uniformity is 0.5, which indicates the zero bits and the one bits are distributed uniformly in PUF response. 
\textbf{Uniqueness} measures the mean Hamming Distance of PUF responses between a pair of different PUF instances (inter-class HD), with the same set of PUF challenges. 
The ideal uniqueness is 0.5, which means each PUF instance produces a unique response distribution. \textbf{Reliability} measures the mean Hamming Distance between actual PUF responses and the enrolled (ideal) PUF responses (intra-class HD). 
The ideal PUF is expected to have no bit flips in responses compared to the enrolled values.

We evaluate the metrics using 1000 random challenges.
Note that during CRP generation, the plaintext $r$ is unbiased since it is sampled from a uniform binary distribution. 
However, when evaluating the lattice PUF response to a particular challenge (ciphertext) $\mathbf{c}$, its actual output $\tilde{r}$ may not match $r$ due to a possibility of decryption error. (Recall from Section \ref{sec:design}, the decryption error rate is targeted at 1.26\%.) 
Therefore, uniformity and uniqueness tests capture the impact of decryption errors on statistical properties of the PUF's true output $\tilde{r}$.

\begin{figure} 
    \centering
  \subfloat[\label{fig:HW}]{%
       \includegraphics[width=0.45\linewidth]{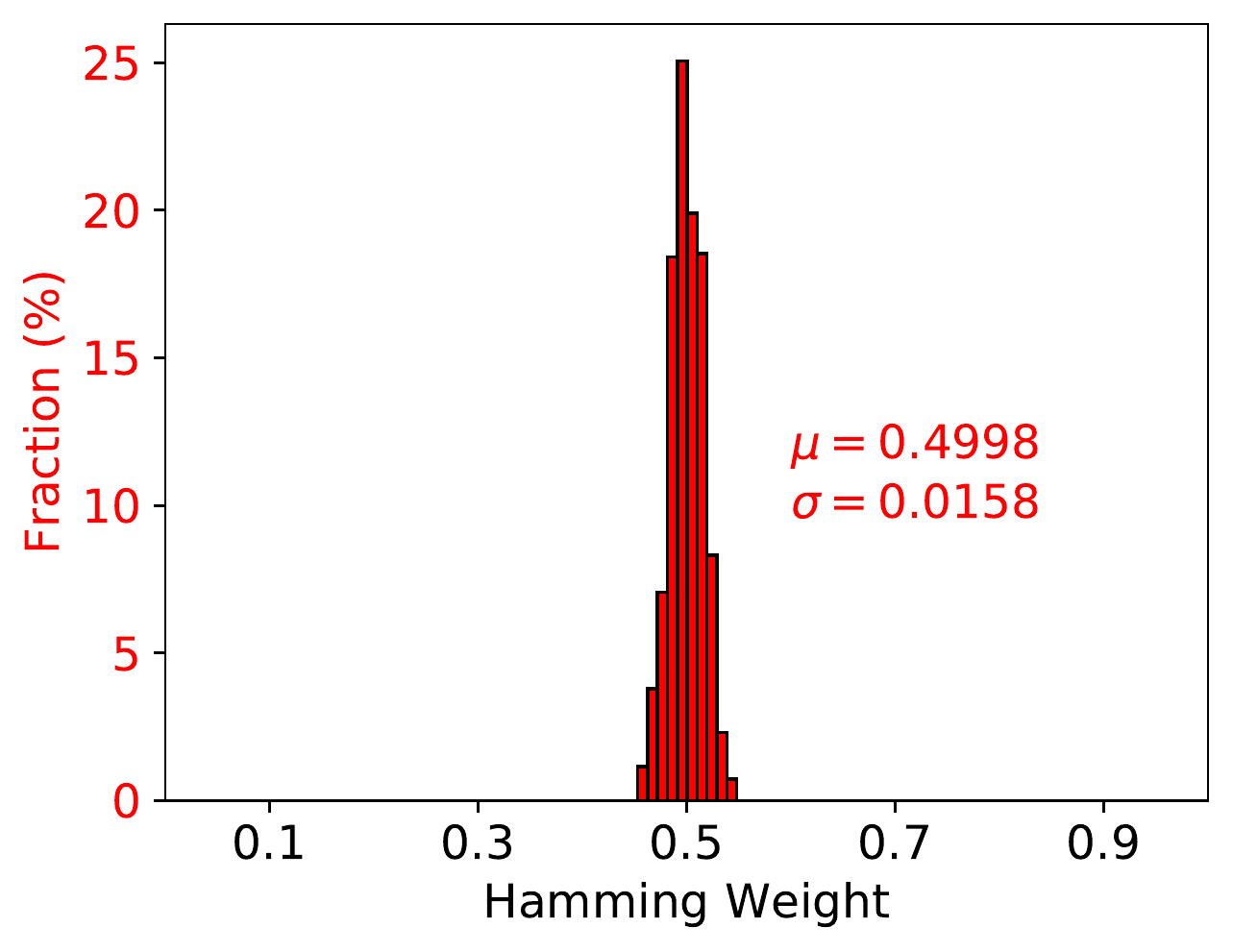}}
    \hfill
  \subfloat[\label{fig:inter_intra}]{%
        \includegraphics[width=0.5\linewidth]{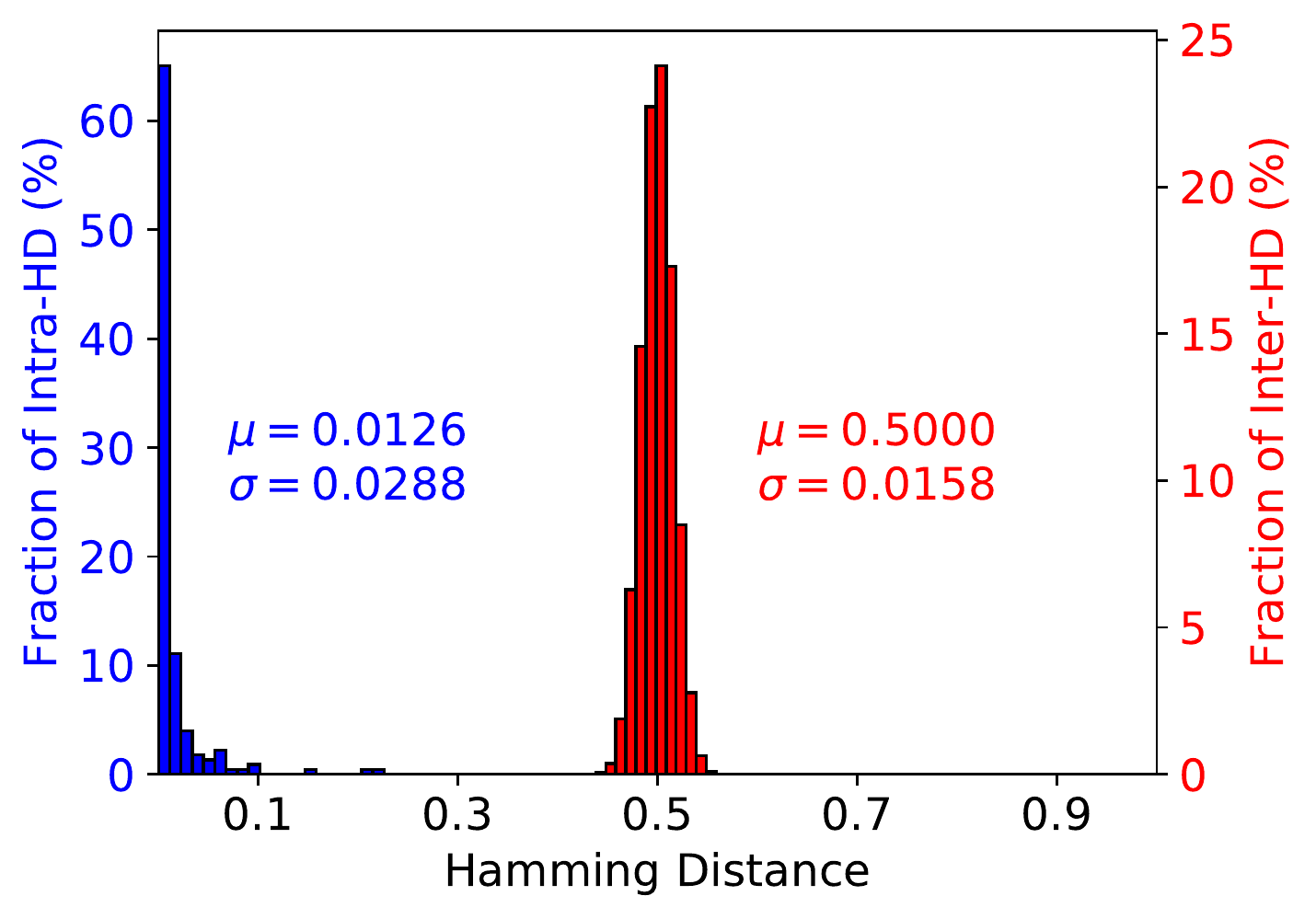}}
  \caption{(a) Lattice PUF has near ideal uniformity. (b) Lattice PUF has near ideal uniqueness and reliability.}
  \label{fig: lattice_puf_stats} 
\end{figure}

The lattice PUF demonstrates near ideal statistical properties: the mean of uniformity distribution is 0.4998 (std = 0.0158) and the mean of uniqueness distribution is 0.5 (std = 0.0158). 
To evaluate PUF reliability, we assume the key reconstruction process is ideal and the errors merely come from LWE decryption. 
Figure \ref{fig:inter_intra} shows the lattice PUF reliability distribution centers close to 0.

\subsection{ML Attacks on Lattice PUF}

Although we establish the security of lattice PUF theoretically, we perform empirical validation as a way to explore the possible attacks of distributional relaxations used, and as a way to give added overall confidence in the design. 
ML algorithms deployed include logistic regression (LR), support vector machine (SVM), one-layer Neural Network (1-NN), and DNN. 

The LR and SVM are implemented with scikit-learn package in Python. 
We use the RBF kernel in SVM, which models a non-linear decision boundary. 
The single hidden layer of 1-NN consists of 100 neurons, and the activation function is selected to be ReLU. 
We train the 1-NN with Keras powered by Tensorflow.

\begin{figure}[t!]
\centering
\includegraphics[width = 0.6\linewidth]{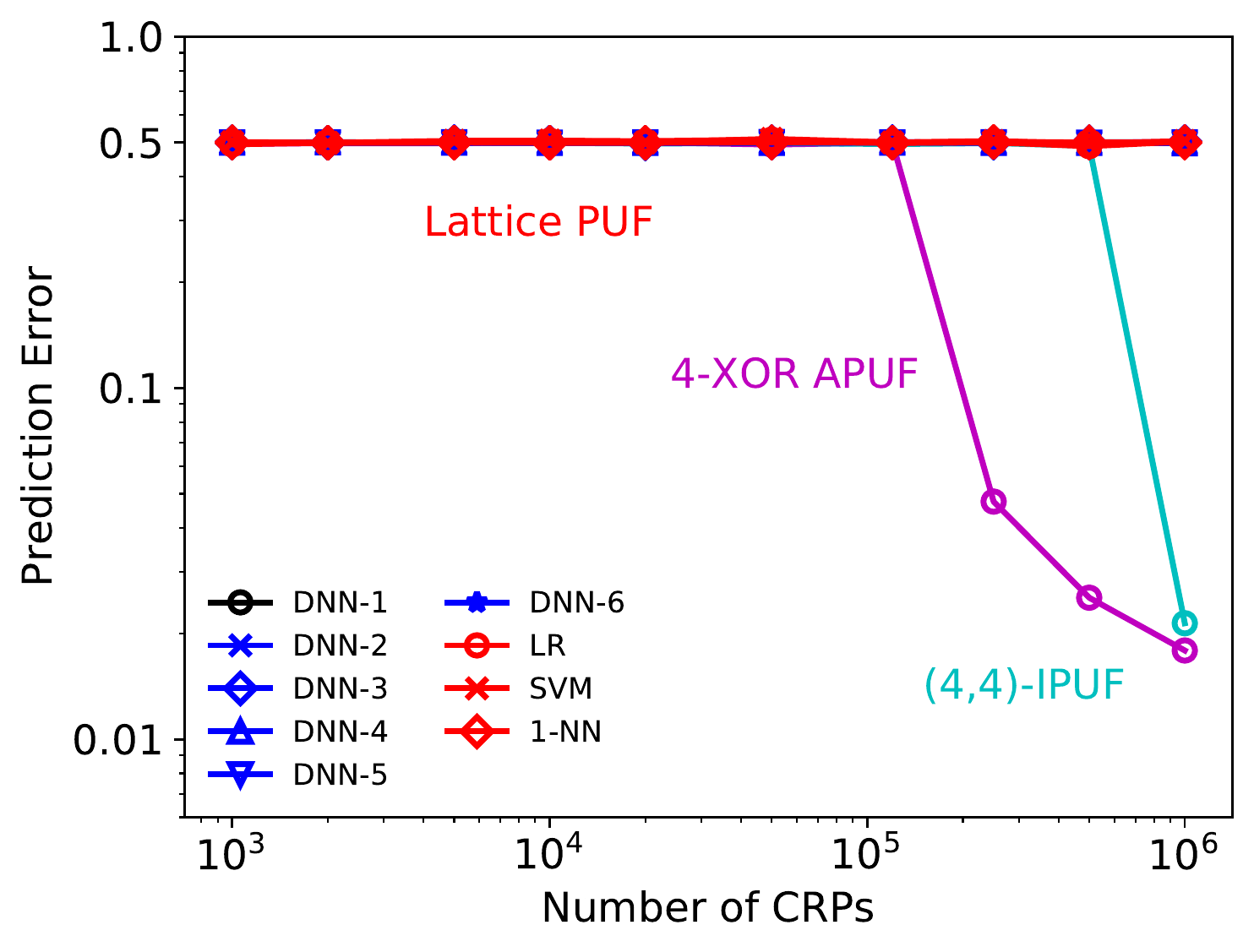}
\caption{Lattice PUF is resilient to multiple ML modeling attacks including DNNs. Two other strong PUFs are ultimately vulnerable to DNN-based attacks.}
\label{fig:ml_attack_2}
\end{figure}

\begin{figure}[t!]
\centering
\includegraphics[width = 0.6\linewidth]{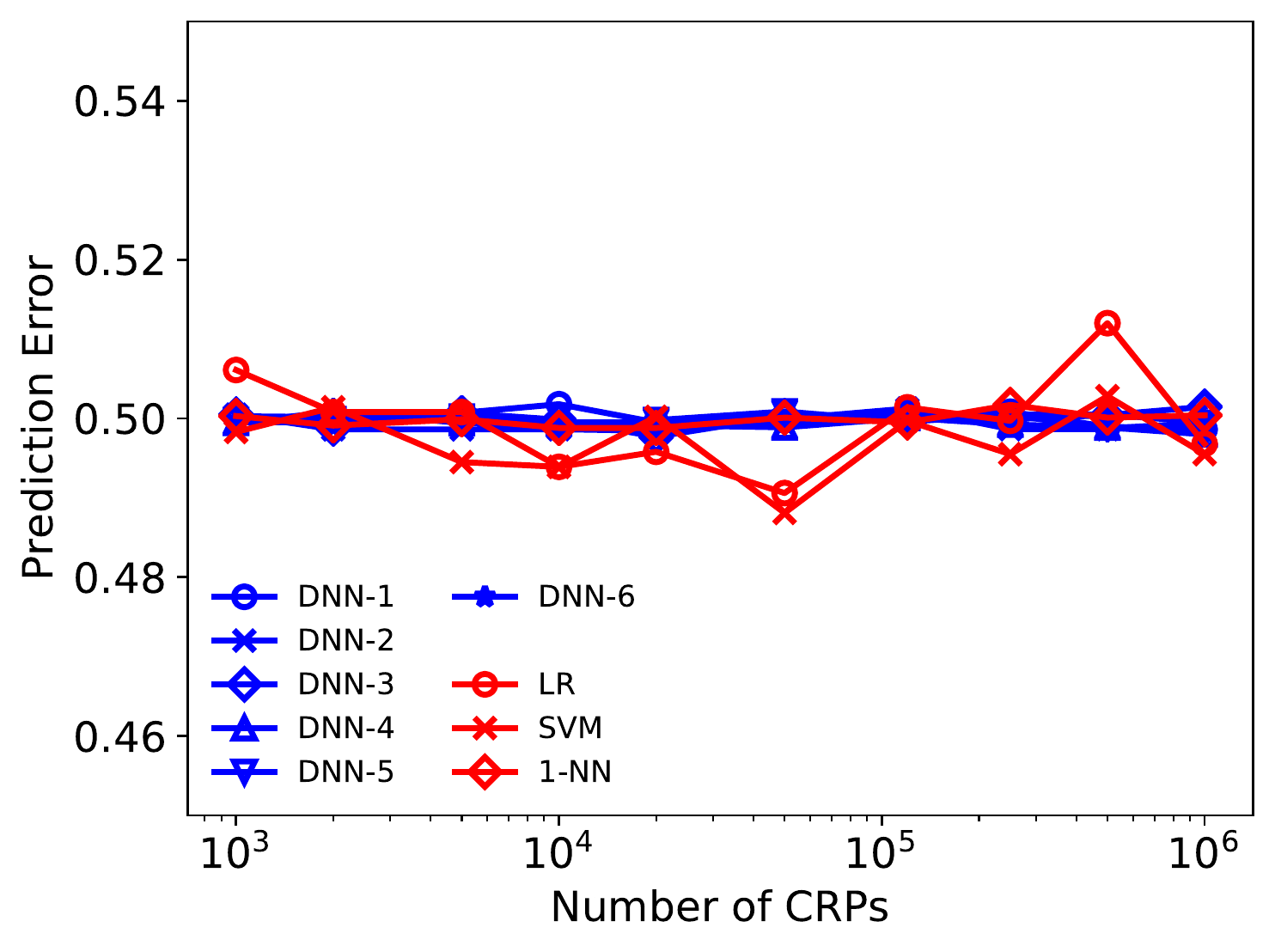}
\caption{Lattice PUF can not be broken by both conventional ML algorithms and more powerful DNNs.}
\label{fig:ml_attack_3}
\end{figure}

\begin{table}[t!]
\centering
	\resizebox{0.9\linewidth}{!}{
        \begin{tabular}{|c|c|c|c|c|c|}
        \hline
        \textbf{Setup} & \begin{tabular}[c]{@{}l@{}}\textbf{Hidden}\\ \textbf{Layers}\end{tabular} & \begin{tabular}[c]{@{}c@{}}\textbf{Neurons}\\ \textbf{per Layer}\end{tabular} & \begin{tabular}[c]{@{}c@{}}\textbf{Challenge} \\ \textbf{Distribution}\end{tabular} & \begin{tabular}[c]{@{}c@{}}\textbf{Input} \\ \textbf{Format}\end{tabular} & \begin{tabular}[c]{@{}c@{}}\textbf{Prediction}\\ \textbf{Error}\end{tabular} \\ \hline
        DNN-1     & 4                                                       & 100                                                         & PRNG                                                              & Binary                                                  & 49.86\%                                                       \\ \hline
        DNN-2     & 4                                                       & 100                                                         & PRNG                                                              & Real                                                    & 49.84\%                                                       \\ \hline
        DNN-3     & 4                                                       & 100                                                         & Ciphertext                                                        & Binary                                                  & 49.76\%                                                       \\ \hline
        DNN-4     & 6                                                       & 100                                                         & PRNG                                                              & Binary                                                  & 49.80\%                                                       \\ \hline
        DNN-5     & 4                                                       & 200                                                         & PRNG                                                              & Binary                                                  & 49.87\%                                                       \\ \hline
        DNN-6     & 12                                                       & 2000                                                         & PRNG                                                              & Binary                                                  &    49.95\%                                                    \\ \hline
        \end{tabular}
    }
    \vspace{1em}
    \caption{Different DNN configurations in modeling attacks.}
    \label{table:DNNSetting}
    \vspace{-2em}
\end{table}

DNN is a class of ML algorithms often highly effective in classification tasks. 
Its superior performance against 1-NNs is enabled by the expressiveness of multiple hidden neuron layers. Notably, recent successful DNN-based attacks have been demonstrated on strong PUFs, such as IPUFs \cite{DBLP:journals/iacr/SantikellurBC19},  XOR APUFs \cite{DBLP:journals/iacr/SantikellurBC19} and XOR BR PUFs \cite{dnn_xor_br_puf}. We select the DNN parameters in \cite{DBLP:journals/iacr/SantikellurBC19} as our baseline configuration (DNN-1): the number of hidden layers is 4, the number of neurons per layer is 100, and the activation function is ReLU. 
Here DNNs are also trained by Keras.

We explored multiple DNN configurations beyond the baseline experiment, Table \ref{table:DNNSetting}. The input to DNN-2 concatenates 161 8-bit integers. The input to DNN-3 follows a different challenge distribution. DNN-4 has an increased number of hidden layers and DNN-5 has an increased number of neurons within each layer. DNN-6 follows the configuration of a recent attack on XOR BR PUFs \cite{dnn_xor_br_puf}, and has a much larger number of both hidden layers and neurons per layer.

The lattice PUF again demonstrates near perfect empirical ML resistance, Figure \ref{fig:ml_attack_2}. The results are reported using a test set with 200K CRPs and the DNNs are trained with the Adam optimizer running for 200 epochs. Regardless of the algorithms and the number of training CRPs used in the attack, the prediction error remains close to 0.5, which is equivalent to a random coin-flip. However, the two other strong PUFs can be successfully modelled by DNNs, with a prediction error smaller than 2\%.

The results also validate the challenge compression technique demonstrated in Section \ref{sec:lfsr}. The DNN-1, DNN-2, DNN-4, DNN-5, and DNN-6 adopt challenges produced by the PRNG, but still show a prediction error close to 0.5. Figure \ref{fig:ml_attack_3} shows a zoomed view of DNN attack results on lattice PUF.

\subsection{Hardware Implementation Results}
\label{sec:hardware_results}

In this section, we show the hardware implementation details of lattice PUF. We synthesize, configure and test the entire design on a low-end XC6SLX45 FPGA device of Xilinx Spartan-6 family.  

We adopt the homogeneous error assumption regarding the FE design. This means all cells are assumed to have the same BER \cite{bosch2008efficient}.
The BER of different POKs varies from $0.1\%$ \cite{karpinskyy20168} to $15\%$ \cite{maes2009soft}.
We configure BER = $1\%$, BER = $5\%$, BER = $10\%$ and BER = $15\%$ to investigate design trade-offs in FE and POK, and finally configure the BER of SRAM POK in our design to be $5\%$. The FE targets a $10^{-6}$ failure rate in reconstruction of $1280$ key bits. 
As described in Section \ref{sec:design}, the overall lattice PUF response BER can achieve the desired value with such a low key reconstruction failure rate.
Our ECC concatenates an inner code and an outer code. The inner code adopts repetition code and the outer code adopts shortened BCH code. Concatenated ECCs usually have shorter code length and lower hardware cost, compared to single codes \cite{bosch2008efficient}.
Table \ref{table:ecc} and \ref{table:hardware_fe} show the parameters and hardware utilization of ECCs with different configurations for BER.
A $5\%$ BER value requires $6.36K$ SRAM cells in order to reconstruct the $1280$-bit secret value $\mathbf{s}$ with a failure rate of $10^{-6}$.
The FE design requires $351$ slices in total.

The lattice PUF (without FE) takes a total of 45 slices on Spartan-6 FPGA. The LFSR and the controller occupies most of the slices. Table \ref{table:fpga_result}(a) shows the resources breakdown of each module. 
The LWEDec block is implemented with MAC unit of 8 bits and a block for MAC result quantization, Figure \ref{fig:lwedec}.
RAM-based shift registers are used to implement the 256-bit LFSR. It takes $47\mu$s in total to generate a single PUF response bit under a clock running at 33.3MHz. 100 PUF response bits require about $4.4ms$ ($8\mu s + 100\times 44\mu s$) in total. Table 
\ref{table:fpga_result}(b) shows latency of each procedure to produce a PUF response.

\begin{table}[t!]
    \centering
    \def\arraystretch{1.1}
    \subfloat[]{
        \resizebox{0.4\linewidth}{!}{
            \begin{tabular}{|c|c|}
            \hline
            \textbf{Module}         & \textbf{Size [slices]} \\ \hline
            LFSR                    & 27            \\ \hline
            LWEDec                  & 2             \\ \hline
            Controller              & 16            \\ \hline
            \textit{Total}          & 45            \\ \hline
            \end{tabular}
        }
    }
    \hspace{1em}
    \subfloat[]{
        \resizebox{0.65\linewidth}{!}{
            \begin{tabular}{|c|c|}
            \hline
            \textbf{Step}                               & \textbf{Time [$\mu$s]} \\ \hline
            Seed $\text{seed}_{\mathbf{a}'}||t$ load for LFSR    & 8             \\ \hline
            1-bit decryption from LWEDec                  & 44            \\ \hline
            \textit{Total} @ 33 \textit{MHz}            & 52            \\ \hline
            \end{tabular}
        }
    }
    \vspace{1em}
    \caption{(a) Area consumption and (b) runtime of our reference lattice PUF implementation on Xilinx Spartan-6 FPGA.}
    \label{table:fpga_result}
\end{table}

\begin{figure}[t!]
\centering
\includegraphics[width = 0.6\linewidth]{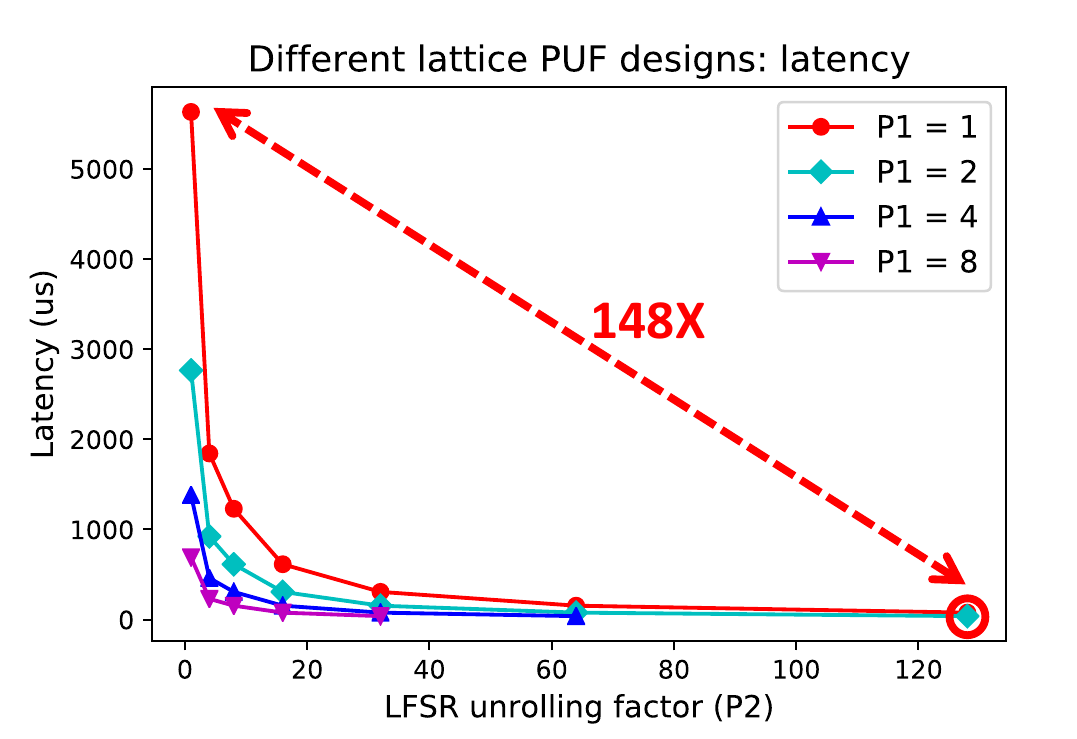}
\caption{Increasing $P_1$ and $P_2$ leads to steady reduction in latency.}
\label{fig:design_space_latency}
\end{figure}

\begin{figure}[t!]
\centering
\includegraphics[width = 0.6\linewidth]{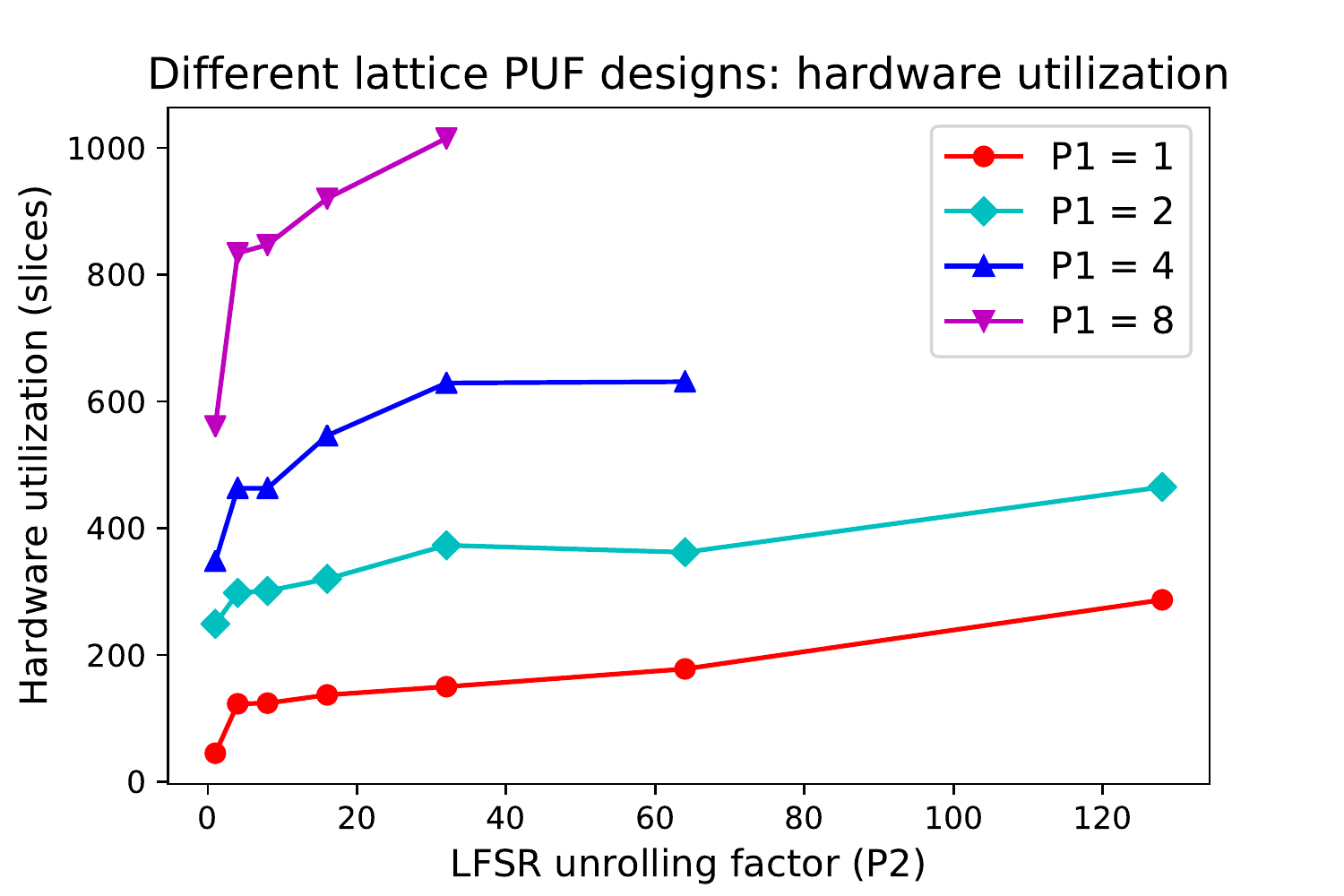}
\caption{Increasing $P_2$ is more hardware efficient than increasing $P_1$, but maximum $P_2$ is limited by the LFSR generator polynomial. The optimal strategy is to increase $P_2$ first to achieve latency goal. If $P_2$ reaches the limit and more aggressive latency reduction is required, increase $P_1$.}
\label{fig:design_space_hw}
\end{figure}

We now explore the design space of our proposed lattice PUF, starting with the resource-efficient design. We adopt the parallelization strategies proposed in Section \ref{sec: lpuf_par} and implement designs with different levels of parallelism. The latency and hardware utilization of the designs are summarized in Table \ref{table:latency_par} and \ref{table:hwslices_par}. Fig. \ref{fig:design_space_latency} and Fig. \ref{fig:design_space_hw} visualize the influence of $P_1$ and $P_2$ on latency and hardware cost. Similar to Table 
\ref{table:fpga_result}(a), we report the sum of slice utilization of LFSR, LWEDec and Controller modules. The maximum value of $P_2$ is $128$ due to the constraint of LFSR generator polynomial. 
Table entries with NA indicate the corresponding design requires more multipliers than the available DSP slices on the device. We observe that the parallelized design leads to a steady reduction in latency, at the cost of increased hardware utilization. We achieve a 148X reduction in latency in the most latency-optimized design, with a 10X increase in hardware utilization. 
In addition, we observe the MAC unit parallelization strategy has better hardware efficiency compared to LFSR-LWEDec datapath parallelization strategy. To achieve the same latency reduction, the design which prioritizes MAC unit parallelization requires fewer slices. For instance, to achieve the optimal $38 \mu$s latency, the design with $(P_1, P_2) = (2, 128)$ only requires $465$ slices, while the design with $(P_1, P_2) = (8, 32)$ requires $1,015$ slices. This validates our optimal strategy demonstrated in Section \ref{sec: lpuf_par}. 

\begin{table}[t!]
\centering
	\vspace{-0.5em}
	\def\arraystretch{1.1}
	\resizebox{0.9\linewidth}{!}{
        \begin{tabular}{|c|c|c|c|c|c|c|}
        \hline
        \textbf{Design} & \textbf{Platform} & \begin{tabular}[c]{@{}c@{}}\textbf{PUF Logic}\\ \textbf{[Slices]}\end{tabular} & \begin{tabular}[c]{@{}c@{}@{}@{}}\textbf{ECC}\\ \textbf{[Slices]}\end{tabular} & \begin{tabular}[c]{@{}c@{}}\textbf{POK}\\ \textbf{[Bits]}\end{tabular} & \begin{tabular}[c]{@{}c@{}}\textbf{Response}\\ \textbf{[Bits]}\end{tabular} & \begin{tabular}[c]{@{}c@{}}\textbf{Latency}\\ \textbf{[$\mu$s]}\end{tabular} \\\hline
        POK+AES \cite{bhargava2014efficient} & Spartan 6 & 80 & 340 & 612 & 128 & 2.2\\ \hline
        Controlled PUF \cite{gassend2008controlled} & Spartan 6 & 127 & 340 & 612 & 256 & 19.1\\ \hline
        \begin{tabular}[c]{@{}c@{}}CFE-based PUF\\ \cite{herder2017trapdoor,jin2017fpga}\end{tabular} & Zynq-7000 & 9,225 & 0 & 450\tablefootnote{The POK is based on RO PUF and assumes a different BER.} & 256 & 658\tablefootnote{Zynq-7000 device typically runs faster than Spartan-6 device.}\\ \hline
        \begin{tabular}[c]{@{}c@{}}Lattice PUF\\ (resource-efficient)\end{tabular} & Spartan 6 & 45 & 351 & 6,360 & 128 & 5,632\\ \hline
        \begin{tabular}[c]{@{}c@{}}Lattice PUF\\ (latency-optimized)\end{tabular} & Spartan 6 & 465 & 351 & 6,360 & 128 & 38\\ \hline
        \end{tabular}
    }
    \vspace{1em}
    \caption{Comparison of hardware utilization of various strong PUFs.}
    \label{table:hardware_puf}
\end{table}

\begin{table}[t!]
    \centering
	\def\arraystretch{1.1}
	\resizebox{0.9\linewidth}{!}{
        \begin{tabular}{|c|c|c|c|c|c|}
        \hline
        \multirow{2}{*}{\begin{tabular}[c]{@{}c@{}}\textbf{Raw BER}\\  \textbf{(\%)}\end{tabular}} & \multicolumn{2}{c|}{\textbf{ECC Configuration}}                                 & \multirow{2}{*}{\textbf{Raw POKs}}  \\ \cline{2-3} 
                                                                                 & Outer code   & Inner code & \\ \hline
        1                                                                        & {[}236, 128, 14{]}  & NA            & 2,360  \\ \hline
        5                                                                        & {[}212, 128, 11{]} & {[}3, 1, 1{]}  & 6,360  \\ \hline
        10                                                                       & {[}220, 128, 12{]} & {[}5, 1, 2{]}  & 11,000  \\ \hline
        15                                                                       & {[}244, 128, 15{]} & {[}7, 1, 3{]}  & 17,080   \\ \hline
        \end{tabular}
    }
    \vspace{1em}
    \caption{Configuration of concatenated ECCs.}
    \label{table:ecc}
\end{table}

\begin{table*}[t!]
    \centering
    \def\arraystretch{1.1}
	\resizebox{0.9\linewidth}{!}{
        \begin{tabular}{|c|c|c|c|c|c|c|c|c|c|}
        \hline
        \multirow{2}{*}{\begin{tabular}[c]{@{}c@{}}\textbf{Raw BER}\\  \textbf{(\%)}\end{tabular}} & \multicolumn{3}{c|}{\textbf{Outer Code}} & \multicolumn{3}{c|}{\textbf{Inner Code}} & \multicolumn{3}{c|}{\textbf{Total}} \\ \cline{2-10} 
                               & Reg      & LUT      & Slice     & Reg      & LUT      & Slice     & Reg    & LUT    & Slice    \\ \hline
        1                      & 899      & 842       & 431        & 0        & 0        & 0         & 899    & 842     & 431       \\ \hline
        5                      & 723       & 650       & 350        & 1        & 2        & 1         & 724     & 652     & 351       \\ \hline
        10                     & 777       & 696       & 386        & 1        & 5        & 2         & 778     & 701     & 388       \\ \hline
        15                     & 963      & 905      & 442        & 1        & 3        & 1         & 964    & 908    & 443       \\ \hline
        \end{tabular}
        }
        \vspace{1em}
        \caption{Hardware utilization in FE design on Spartan 6 FPGA.}
    \label{table:hardware_fe}
\end{table*}

\begin{table}[t!]
\centering
\def\arraystretch{1.1}
	\resizebox{0.9\linewidth}{!}{
\begin{tabular}{|c|*{7}{c|}}\hline
\backslashbox{\textbf{$\mathbf{P_1}$}}{\textbf{$\mathbf{P_2}$}}
&\makebox{\textbf{1}}&\makebox{\textbf{4}}&\makebox{\textbf{8}}
&\makebox{\textbf{16}}&\makebox{\textbf{32}}&\makebox{\textbf{64}}&\makebox{\textbf{128}}\\\hline
\textbf{1} & 5,632 & 1,843 & 1,229 & 614 & 307 & 154 & 77\\\hline
\textbf{2} & 2,765 & 922 & 614 & 307 & 154 & 77 & 38\\\hline
\textbf{4} & 1,382 & 461 & 307 & 154 & 77 & 38 & NA\\\hline
\textbf{8} & 691 & 230 & 154 & 77 & 38 & NA & NA\\\hline
\end{tabular}
}
\vspace{1em}
\caption{Latency of different lattice PUF designs for 128-bit response generation [$\mu$s]. $P_1$ denotes number of LFSR-LWEDec data-paths; $P_2$ denotes number of LFSR output bits.}
\label{table:latency_par}
\end{table}

\begin{table}[t!]
\centering
\def\arraystretch{1.1}
	\resizebox{0.9\linewidth}{!}{
\begin{tabular}{|c|*{7}{c|}}\hline
\backslashbox{\textbf{$\mathbf{P_1}$}}{\textbf{$\mathbf{P_2}$}}
&\makebox{\textbf{1}}&\makebox{\textbf{4}}&\makebox{\textbf{8}}
&\makebox{\textbf{16}}&\makebox{\textbf{32}}&\makebox{\textbf{64}}&\makebox{\textbf{128}}\\\hline
\textbf{1} & 45 & 123 & 124 & 137 & 150 & 178 & 287\\\hline
\textbf{2} & 249 & 298 & 301 & 320 & 373 & 362 & 465\\\hline
\textbf{4} & 348 & 463 & 463 & 546 & 629 & 631 & NA\\\hline
\textbf{8} & 561 & 834 & 847 & 920 & 1015 & NA & NA\\\hline
\end{tabular}
}
\vspace{1em}
\caption{Hardware utilization of different lattice PUF designs [slices]. $P_1$ denotes number of LFSR-LWEDec data-paths; $P_2$ denotes number of LFSR output bits.}
\label{table:hwslices_par}
\end{table}

We finally compare the hardware utilization and latency of lattice PUF designs to several published strong PUF designs \cite{bhargava2014efficient, gassend2008controlled, jin2017fpga} which also have ML resistance in Table \ref{table:hardware_puf}. 
As mentioned in Section \ref{sec:intro}, there are two different categories of ML resistance among those listed PUFs. 
ML resistance of the AES-based PUF and the controlled PUF can be reduced to the security of the deployed primitives. 
In contrast, the theoretical ML resistance of the CFE-based strong PUF and our lattice PUF can be reduced to the hardness of fundamental math problems, like LPN and LWE.
The original proposal of the AES-based strong PUF \cite{bhargava2014efficient} is an ASIC implementation. 
Here, to estimate the AES implementation cost in FPGA, we adopt results in \cite{chu2012low}.
Note that the original proposal of \cite{bhargava2014efficient} does not use FE-based error correction: it uses dark bit masking to guarantee reliability. 
To estimate the cost of its error correction on an FPGA implementation, we use the FE design for a 128-bit key reconstruction with a $5\%$ raw BER and $10^{-6}$ failure rate, similar to the ECC design for the lattice PUF. The estimated number of raw POK bits are derived using these parameters.
Similarly, to calculate hash function hardware utilization in the controlled PUF \cite{gassend2008controlled}, the FPGA implementation results of SHA-3 in \cite{sha3_finalist} is used. 
No design details of the ECC and the POK are given in the proposal of controlled PUF \cite{gassend2008controlled}. 
We follow the same assumption as for the AES PUF for estimation.
\cite{jin2017fpga} reports the hardware utilization results of the CFE-based strong PUF on FPGA in the LUT numbers.
By utilizing \cite{xilinx:ds190}, we convert the LUT numbers to the slice numbers. 
The CFE-based strong PUF does not use a conventional FE-based error correction so its ECC cost is 0.

Table \ref{table:hardware_puf} shows that the resource-efficient version of the lattice PUF achieves the smallest slice number among all strong PUF designs with ML resistance.
From the latency perspective, the AES-based PUF and the controlled PUF run faster than the CFE-based PUF and our lattice PUF. This is primarily due to the fundamental algorithmic difference of the LWE problem and the established cryptographic primitives. The LWE problem requires sequential MACs on the elements of long vectors and produces a single output bit per execution, while AES finishes within a relatively small number of rounds and has a much larger output size.
Compared to the CFE-based strong PUF whose ML resistance can also be reduced to the hardness of a fundamental math problem, the resource-efficient lattice PUF achieves a 205X area reduction, and the latency-optimized version achieves a 16X latency improvement together with a 19X area reduction.

\section{Conclusion}
\label{sec:conclusion}
\vspace{-1mm}
We propose a novel strong PUF with theoretically proven security against ML attacks conducted by both classical and quantum computers. 
Its security is guaranteed by the cryptographic hardness to learn decryption functions of public key cryptosystems. 
Our PUF is constructed from the LWE decryption function. 
A series of designs are implemented on a Xilinx Spartan-6 FPGA. 
A compact design uses a highly serialized LFSR and LWEDec block, while a latency-optimized design uses an unrolled LFSR and a parallel datapath. 
The lattice PUF designs have a CRP space of $2^{136}$, with $6,360$ POK bits and $128$-bit concrete ML resistance. 
Excellent statistical characteristics are demonstrated.   
\section*{Acknowledgments}
We thank Dr.~Aydin Aysu for his insightful advice on idea presentation, assistance with FPGA implementation of repetition code, and comments that greatly improved the manuscript.

\bibliographystyle{abbrv}
\bibliography{main.bib}

\end{document}